\documentclass[reprint,
superscriptaddress,aps,pre,twocolumn,floatfix,showpacs,amsmath,10pt]{revtex4-1}

\usepackage{booktabs}
\usepackage{xcolor}
\usepackage{graphicx}
\usepackage{graphics}
\usepackage{amsmath}
\usepackage{amssymb}
\usepackage{amsthm,amstext}
\usepackage{amsfonts}
\usepackage{lipsum}
\usepackage{MnSymbol}
\usepackage{mathrsfs}
\usepackage{stmaryrd}
\usepackage{latexsym}
\usepackage[applemac]{inputenc}
\usepackage{accents}

\usepackage[bbgreekl]{mathbbol}
\DeclareSymbolFontAlphabet{\mathbbm}{bbold}
\DeclareSymbolFontAlphabet{\mathbb}{AMSb}
\DeclareMathAlphabet\mathbfcal{OMS}{cmsy}{b}{n}

\newcommand\ba{\textbf{\emph{a}}}

\newcommand\be{\textbf{\emph{e}}}

\newcommand\by{\textbf{\emph{y}}}

\newcommand\bu{\textbf{\emph{u}}}
\newcommand\bz{\textbf{\emph{z}}}
\renewcommand\bm{\textbf{\emph{m}}}
\newcommand\bx{\textbf{\emph{x}}}

\newcommand\br{\textbf{\emph{r}}}
\newcommand\bs{\textbf{\emph{s}}}
\newcommand\bn{\textbf{\emph{n}}}

\newcommand\bk{\textbf{\emph{k}}}

\newcommand\f{\textbf{\emph{f}}}
\newcommand\g{\textbf{\emph{g}}}

\newcommand\0{\textbf{\emph{0}}}

\newcommand\F{\textbf{F}}

\newcommand\Q{\textbf{Q}}
\newcommand\R{\textbf{R}}
\newcommand\G{\textbf{G}}

\renewcommand\S{\textbf{S}}

\newcommand\A{\textbf{A}}
\newcommand\I{\textbf{I}}

\renewcommand\d\delta
\newcommand\D\Delta
\newcommand\e{\varepsilon}
\newcommand\s{\sigma}
\newcommand\ph{\varphi}

\newcommand\bsigma{\boldsymbol{\sigma}}

\newcommand\bet{\boldsymbol{\eta}}
\newcommand\bpsi{\boldsymbol{\psi}}
\newcommand\beps{\boldsymbol{\epsilon}}

\newcommand\sym{\text{sym}}

\newcommand\Curl{\text{Curl}}

\newcommand\Div{\text{Div}}

\newcommand\tr{\text{tr}}

\newcommand\Body{\mathcal{B}}

\newcommand\bbP{\mathbb{P}}
\newcommand\bbC{\mathbb{C}}

\newcommand\bbR{\mathbb{R}}

\newcommand\beq{\begin{equation}}
\newcommand\beqn{\begin{eqnarray}}
\newcommand\eeq{\end{equation}}
\newcommand\eeqn{\end{eqnarray}}

\newcommand{\trsp}{^{\hspace{-1pt}\textsf{T}\hspace{-1pt}}}

\usepackage[normalem]{ulem}

\begin{document}

\title{Nonlinear elasticity of incompatible surface growth}

\author{Lev Truskinovsky}
\email{trusk@lms.polytechnique.fr}
\affiliation{PMMH, CNRS UMR 7636, PSL, ESPCI,
10 rue de Vauquelin, 75231 Paris, France.}

\author{Giuseppe Zurlo}
\email{giuseppe.zurlo@nuigalway.ie}
\affiliation{ School of Mathematics, Statistics and Applied Mathematics, NUI Galway, University Road, Galway, Ireland.}

\begin{abstract}
Surface growth is a crucial component  of many natural and artificial  processes from cell proliferation to additive manufacturing. In elastic systems surface growth is usually accompanied by the development of geometrical incompatibility leading to residual stresses and triggering various instabilities. In a recent paper (PRL, 119, 048001, 2017) we  developed a linearized elasticity theory of incompatible surface growth which  quantitatively linked  deposition protocols with   post-growth states of stress.  Here we  extend this analysis to account for both physical and geometrical nonlinearities of an elastic solid. The new development reveals  the  shortcomings of the linearized theory, in particular,   its inability to describe kinematically confined surface growth and  to account for  growth-induced elastic instabilities.
\end{abstract}

\maketitle


\section{Introduction}

A variety of natural and artificial processes rely on  active mass deposition on the surface of a solid body. The associated class of phenomena is quite broad including such diverse processes as growth of plants \cite{Archer,Dumais2001}, cell motility \cite{Dafalias,johnpre14}, construction of retaining walls \cite{GoodmanSlopes,Labuz}, formation of planets \cite{Kadish2005}, crystallisation from solution \cite{Wildeman,Fink,Schwerdtfeger} and 3D printing  \cite{Ge,LindNature16}. Surface growth is understood in this context  as a  continuous addition of new layers of mass on the external boundary of a solid. From the perspective  of  elasticity theory,  the  fundamental interest here is in  the fact that the accreeted  mass points  arrive with their own \emph{reference state},  which must \emph{emerge} as an outcome  of the manufacturing process. 

Despite considerable  interest  of \emph{compatible} surface growth  \cite{Skalak, Epstein, DiCarloSurf, MauginCiarletta, Ganghoffer2010, Tomassetti, MoultonGoriely, Jenkins}, here we focus on  the case when the manufactured  reference state is \emph{ incompatible} (non-Euclidean) in the sense that it cannot be realized in 3D without generating residual stresses.   The underlying ``geometric frustration'' \cite{kondo49,Efrati,RazPNAS,Sharon},  which  is  ultimately \emph{shaped} by the deposition process,  may be   beneficial (as in growing plants \cite{GorielyPlants})  or detrimental (as in civil  engineering  structures \cite{Hossain}), as it was already exemplified in the  early attempts to understand  incompatible surface growth  motivated by the necessity to explain   the built-up of ``growth stresses'' in trees \cite{Martley}, to  optimize the  concrete pouring \cite{Arutyunyan}   and to improve the quality of  industrial winding  \cite{Southwell}.   

The first systematic theoretical study of the effects of incompatibility in surface growth was conducted by the  Russian school \cite{Rashba, Trincher, Naumov,Manz14,Drozdov98,Lychev17} with the largely parallel development and subsequent extension of the theory in the West \cite{Goodman,Fletcher,Zabaras,Gambarotta,SozioYavari,GuptaTrees,GanghofferGoda,Papadopoulos,Goriely19,Bulging}. In particular, these studies  have raised an awareness  of the presence of a ``historical element'' in the incompatible surface growth problems, which  implies that accumulated  inelastic strains   keep a detailed  \emph{memory} of the deposition process. In defiance of these efforts, however, the  relation between the accretion protocol and the ensuing state of geometric frustration remains poorly understood. The problem is that this relation is inherently \emph{nonlocal} in both, space and time due to the long range character of elastic interactions, and  the unavoidable coupling between  the incremental adjustments of the elastic state to the advances of the accretion front. 
 
This problem was addressed in the two recent papers \cite{ZTprl,ZTMaugin} where we  developed a linearized theory of   incompatible  surface growth focused on the deposition-protocol-dependence of the resulting state of residual stress. The analysis in the linear case turned out to be relatively simple, primarily, because the geometry could be  decoupled from elasticity.  In the present paper we generalize this theory  by taking into account  both geometrical (e.g. finite rotations) and physical (e.g. finite stretches) nonlinearities of an elastic solid. 

Geometric nonlinearities are particularly relevant when the kinematics of the growth process cannot be linearized, as, for instance,  in the case of actin polymerization against a solid wall \cite{johnpre14,Misbah2016}, where  the identification of the reference and the actual states  is incompatible with the very presence of the incoming  mass flux.  Physical  nonlinearities are crucial, for instance,  when due to extreme dependence of the elastic moduli on pre-stress \cite{MacKintosh},  growth induced  inhomogeneity  can  lead to dramatic spatial heterogeneity of the elastic response.  The use of a nonlinear theory becomes, of course,   imperative when one deals with soft solids like   biological tissues or  synthetic gels  \cite{Nardinocchi,Teresi}.

In this paper we formulate a general theory of large-strain incompatible surface growth. To illustrate the specific effects of nonlinearity we present  a systematic analysis of the  case of spherical symmetry, where partial differential equations reduce to ordinary differential equations and at least some of the computations can be performed analytically. Several case studies  were chosen to show the details of the implementation of the  general approach. 

Our first example deals with winding protocols  producing disclination-type incompatibility. The second example shows how such singular incompatibility   reveals itself if  a 2D   disk   is allowed to relax into the 3D space. The third example illustrates  mass accretion under prescribed external pressure, emphasizing  extreme sensitivity of the  embedded incompatibility to  small changes in the deposition protocol.   In the  last two examples we consider outward and inward surface growth against rigid constraints. Such problems are out of reach for  geometrically linearized theory, moreover,  we show that  the account of physical nonlinearities in this context allows one to predict the emergence of  growth-induced   material instabilities.
  
 The paper is organized as follows. In Sec. \ref{Sec2} we develop the concept of inelastic  surface growth and discuss the geometrical meaning of incompatibility. In Sec. \ref{Sec3} we compare various deposition protocols and formulate the corresponding conditions on the growth surface. In Sec.\ref{Sec4} we present the incremental formulation of the accretion problem in the general nonlinear setting. In Sec. \ref{Sec5} we use the nonlinear theory to obtain its  linearized version developed in \cite{ZTprl,ZTMaugin}. The specialization of the general theory for the case of spherical symmetry is discussed in Section Sec. \ref{Sec6}. The case studies, illustrating various effects of physical and geometrical nonlinearity,  are presented in Sec. \ref{Sec7}.  Finally, Sec. \ref{Sec8} contains our conclusions.

\section{Preliminaries\label{Sec2}}

 Consider a 3D body $\mathring\Body$  equipped with an arbitrary set of (Lagrangian) coordinates $\bx$ and  a metric tensor $\G(\bx)$  allowing one to  measure distances in $\mathring\Body$.  Suppose that  $\G$ is compatible  in the sense  that   
there exists  the deformation (embedding) of $\mathring\Body$  in the Eucledian space $\Body=\by(\mathring\Body)$ such that  $\G=\nabla\by\trsp\nabla\by$ where $\by=\by(\bx)$.  Suppose  also  that $\mathring\Body$ is   equipped with a reference metric $\mathring\G(\bx)$ which does not have to be compatible and may carry  information about  the local configuration of defects.

The elastic response is calibrated by  the  distance between $\G$ and $\mathring\G$, so we can   introduce an elastic energy density $e(\G,\mathring\G) \geq 0$, such that $e(\mathring\G,\mathring\G)=0$. In equilibrium  the (Piola-Kirchhoff) stress tensor  $\S =\partial e/\partial \nabla\by$ must satisfy the  conditions 
\beq\label{equires1} 
\left\{
\begin{array}{lll}
\Div\S(\bx) + \f(\bx)  = \textbf{0} & &  \mathring\Body\\
\S(\bx) \mathring\bn(\bx)   = \bs(\bx)  & &  \partial \mathring\Body
\end{array}
\right.
\eeq
 where $\mathring\bn$ is the normal to $\partial\mathring\Body$. We  assumed here that the body $\mathring\Body$  is loaded by surface tractions  $\bs(\bx)$ and body forces $\f(\bx)$. 
 
 \begin{figure}[thb] 
\includegraphics[width=7.5cm]{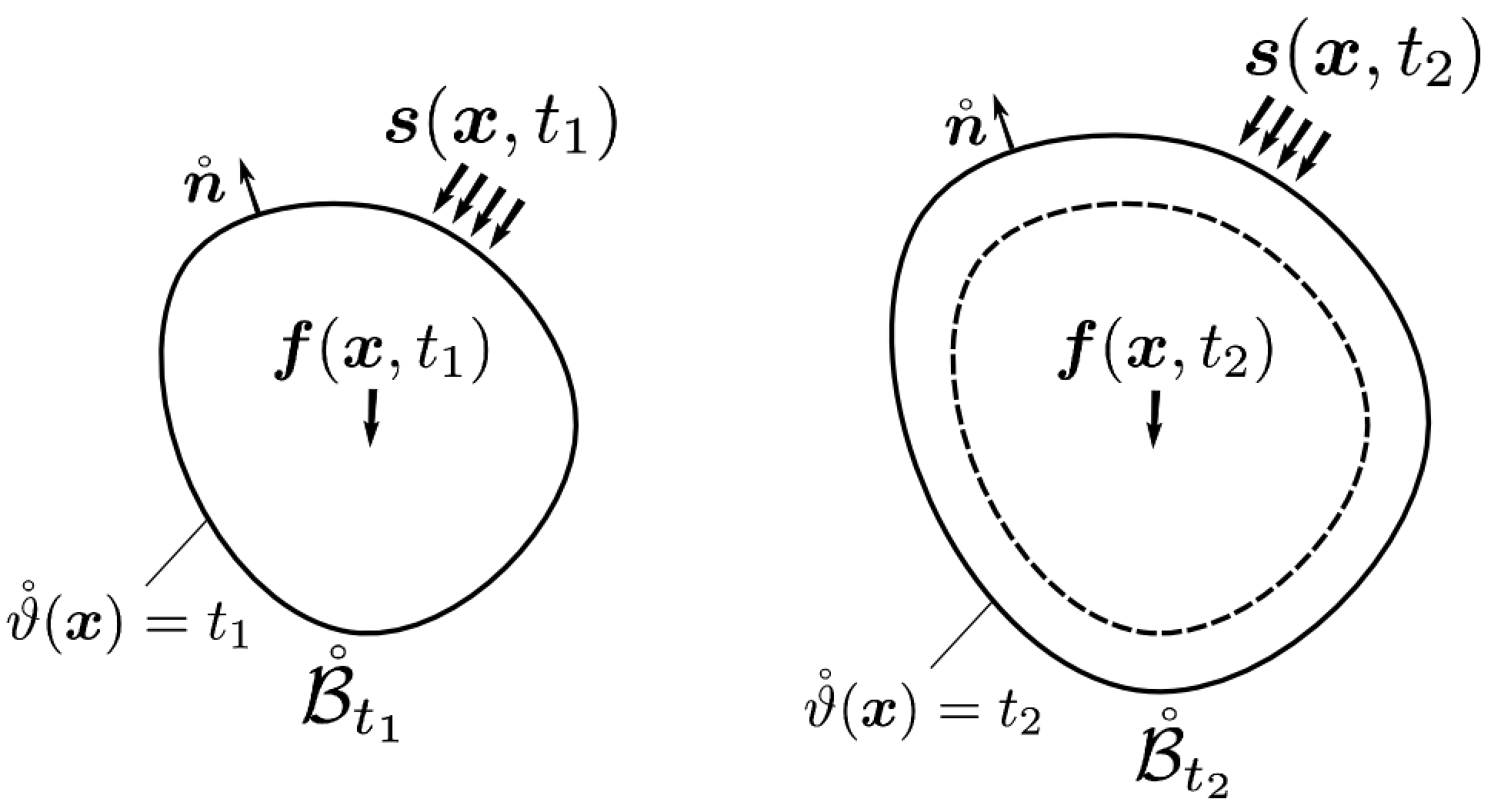}
\caption{\label{sequence} A schematic representation of  two successive reference configurations for the growing body with  $t_1<t_2$.   The configuration  corresponding to $t=t_1$ is shown by a  dashed line in the snapshot corresponding to $t=t_2$.}
\end{figure}

Suppose now that the body $\mathring\Body_t$ is growing due to  addition of points on the evolving boundary $\partial\mathring\Body_t$, see Fig.\ref{sequence}. This process can be described by prescribing a function $\mathring\vartheta(\bx)=t$ that defines the placement of the referential surface at a time instant $t$. The normal to this boundary is $\mathring\bn=||\nabla\mathring\vartheta ||^{-1} \nabla\mathring \vartheta $ and its normal velocity is $\mathring D=||\nabla\mathring\vartheta ||^{-1} $. Given that inertial terms can be neglected, the  deformation of the growing body $\by(\bx,t)$ must satisfy a one-parametric family of equilibrium equations
\beq\label{equires2} 
\left\{
\begin{array}{lll}
\Div\S(\bx,t) + \f(\bx,t)  = \textbf{0} & &  \mathring\Body_t\\
\S(\bx,t) \mathring\bn(\bx,t)   = \bs(\bx,t)  & &  \partial \mathring\Body_t
\end{array}
\right. 
\eeq
where  $\bs(\bx,t)$ and $\f(\bx,t)$  can be now time dependent. If the reference metric $\mathring\G (\bx,t)$ is   known and  the energy density $e$ satisfies suitable conditions \cite{Ball}, the system \eqref{equires2} can be solved at each instant. 

To simplify the problem we can assume  that the reference metric $\mathring\G(\bx)$ is time independent, which means that it is fixed for each material particle at the stage of deposition and is not evolving afterwords. The challenge is to link   physically realizable ``controls''  on the evolving surface  with  particular  distributions of $\mathring\G(\bx)$ in the grown body \cite{Vanel}.  

The   reference metric $\mathring\G$ carries   information about \emph{how} the body is assembled, which  is contained in the Ricci tensor \textbf{R}$(\bx,\by)=\tsum_{i}\langle \mathcal{R} (\bx,\be_i)\by,\be_i\rangle$, a contraction of the Riemann-Christoffel tensor $ \mathcal{R} (\bx,\by)\bz=\nabla_{\bx}\nabla_{\by}\bz - \nabla_{\by}\nabla_{\bx}\bz - \nabla_{[\bx,\by]}\bz$, where $\bx,\by,\bz$ are arbitrary vectors,   $\left\{\be_i\right\}$ is an ortho-normal basis in $\mathbb{R}^3$ and $\nabla$ the Levi-Civita connection induced by $\mathring\G$ (see \cite{Kuhnel} for details).

If   $\R(\mathring\G) \equiv 0$ there  exists   a map $\mathring\g$ such that $\mathring\G=\nabla\mathring\g\trsp\nabla\mathring\g$  \cite{Ciarlet} and  the body is unstressed in the absence of loading ($\bs=\0,\f=\0$). Indeed, the equilibrium  equations in this case 
\beq\label{equires3}
\left\{
\begin{array}{lll}
\Div\S(\bx,t) =0 && \mathring\Body_t\\
\S (\bx,t) \mathring\bn (\bx,t)=0 &&  \partial\mathring\Body_t. 
\end{array}
\right.
\eeq
have a  homogeneous solution $\S(\nabla\by,\nabla\mathring\g\trsp\nabla\mathring\g)=\textbf{0}$   with $\by\equiv \mathring\g$. Such  growth is compatible and the map $\mathring\g$ only affects the stress-free shape of the  body $\mathring\g(\mathring\Body_t)$, see for instance \cite{Nardinocchi}. 

 If, instead,  $\R(\mathring\G)\neq 0$, then it is not possible to find a smooth deformation $\by$ such that $\S(\nabla\by,\mathring\G)=\textbf{0}$ and the unloaded body is   prestressed. Since the tensor \textbf{R}$(\mathring\G)$ identically satisfies a set of 3 scalar differential constraints (Bianchi identities),  $\Div(\textbf{R} - S\,\mathring\G/2)=0$, where  $S=\tsum_{i,j}\langle\mathcal{R} (\be_i,\be_j)\be_j,\be_i\rangle$, only three independent components  of $\R(\mathring\G)$ characterize  the  distribution of growth related  defects \cite{Davini, RazPNAS, Hsu68, Teresi}.  The presence of such defects is a sign of geometric frustration, which can be viewed as an embedded  ``information", revealed  through residual stresses.

\section{Deposition protocols\label{Sec3}}

Suppose that the presence of an active agent on the growing surface can be modeled  by  \emph{nonstandard}  boundary conditions defining the ``growth protocol". One possibility is to directly prescribe the functions $\mathring\G (\bx)$  on the growth surface. Three of its ``incompatible" components will then describe the   accumulated defects and  acquired residual stresses,  while the other three ``compatible" components will characterize  the shape of the  body at the end of the deposition process; the latter  is relevant in many   biological problems \cite{Nardinocchi}  and in engineering problems related to residual  actuation  \cite{Danescu2018}.  

However, the case when  the functions $\mathring\G (\bx,t)$ can be directly controlled  on the growth surface  is not very realistic.  Instead, the reference state is ``manufactured''  during deposition, as a result of the physical actions that one can control through the deposition machinery. Typical  examples of natural processes where $\mathring\G (\bx)$ is not fixed a priori,  include layered manufacturing \cite{Gambarotta},  3D printing \cite{Ge} and biological accretion \cite{Skalak,Skalak82}.

\subsection{Stress control}

In view of the mechanical nature of  an ``active agent'' involved in the execution of a deposition protocol, it is natural to formulate the  corresponding boundary conditions in terms of  the deposition  stress $\overline{\S}(\bx) =\S(\bx,\mathring\vartheta(\bx))$. Here and in what follows, the overline bar will indicate the restriction of a function to $\partial\mathring\Body_t$ \cite{NoteBar}. 

The tensor function $\overline{\S}(\bx)$ is a priori partially constrained by the applied tractions  $\overline{\S}\mathring\bn=\bs$ and by the fact that the angular momentum is balanced, $\overline{\S}(\overline{\nabla\by})\trsp=(\overline{\nabla\by})\overline{\S}\,\trsp$.  This implies that out of nine scalar components of $\overline{\S}$, six are essentially known. The remaining three can  be still  used to at least partially constrain the distribution of embedded inhomogeneity, and will constitute our first set of \emph{active  controls}. 

\begin{figure}[thb] 
\includegraphics[width=6.5cm]{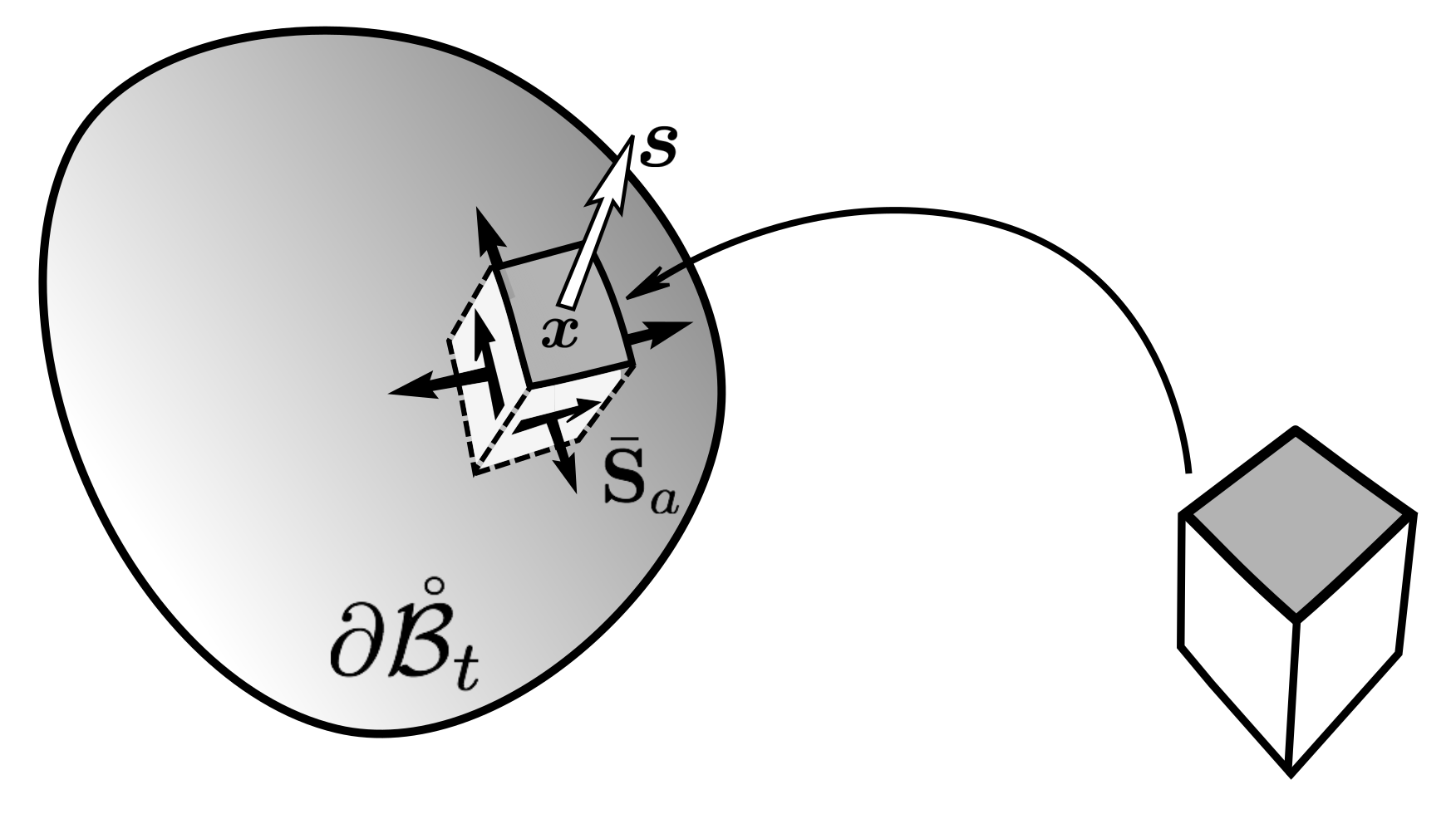}
\caption{\label{gensch} Schematic representation of the deposition protocol for an arriving unloaded ``brick''  which is  placed on the   growth surface  after being subjected to  ``passive'' tractions $\bs=\S\mathring\bn$ and   ``active'' surface stresses  $\bar{\S}_a$.}
\end{figure}

To be more specific, consider the decomposition of the deposition stress   
\beq
\overline{\S}(\bx) =  \S_a(\bx) + \S_p(\bx) \qquad \bx\in\partial\mathring\Body_t
\eeq
where the ``passive'' contribution   $\S_p(\bx)$ is  defined by the applied surface tractions  $\S_p=\bs\otimes\mathring\bn+\mathring\bn\otimes\bs -(\bs\cdot\mathring\bn)\mathring\bn\otimes\mathring\bn$ and the ``active''  component is restricted to the surface $\S_a\mathring\bn=0$, see Fig.\ref{gensch}. Fixing the  three independent components of  $\S_a(\bx)$ can be then considered as a stress-control part of the deposition protocol. 

As we show in our examples, the stress control can be performed not only through the Piola stress $ \S(\bx)$ but also through the Cauchy stress $\bsigma(\by)$. The two are related through $\S=\bsigma\text{cof}(\nabla\by)$, where $[\text{cof} (\A)] \trsp = \A^{-1} \det\A $,  and the difference between them is an effect of geometrical  nonlinearity.

\subsection{Kinematic control}
 
The other three controls can be, for instance,  of kinematic origin and involve  constraints on the current position of the growing surface, which amounts to  prescribing the functions $\bar\by(\bx) = \by(\bx,\mathring\vartheta(\bx))$.  Such constraints can be imposed passively as, in the case of growth against a fixed wall, or actively, as in the case of growth under an oscillating piston. 

Another important case when the  controls may involve  the field $\bar{\by}(\bx)$ is  that  of solidification, where the built-up of pressure at the growth interface depends on the current position of the solidification front \cite{Fletcher}. In those cases one can  use the fact that  the map $\bar\by$ completely characterizes the    areal stretch $ \alpha= ||\text{cof}\,(\nabla\bar\by)\mathring\bn||$ and the normal $\bn=\alpha^{-1}\text{cof}\,(\nabla\bar\by)\mathring\bn$ of the current surface  $\partial\Body_t$. 

Note that the control of   the actual position of the growth surface  $ \by(\bx,\mathring\vartheta(\bx))$ implies the knowledge  of   the reference position of this surface defined by the function  $\mathring\vartheta(\bx)$. The knowledge of  the latter is equivalent to fixing  the rate of  mass delivery  at the   referential  growth surface   $\dot m=\mathring{\varrho}\mathring{D}$, where $\mathring{\varrho}$ is the referential density.

 We can also express the arriving mass flux in terms of the actual parameters {\color{red}}  $\dot m =\bar{\varrho}\alpha (D-\bar{\dot{\by}}\cdot\bn) $, where $\bar{\varrho}=\mathring{\varrho} (\det \overline{\nabla \by})^{-1}$ is the actual density     and $D=\mathring{D}(\nabla\bar{\by})\mathring{\bn}\cdot\bn$ is the Eulerian velocity of the growing surface. Using the (commutation) relation 
 \beq\label{com}
\nabla \bar \by -\overline{\nabla\by} \, = \,  ||\nabla\mathring\vartheta|| {\overline{\dot \by}}\otimes\mathring \bn
\eeq
we may rewrite the expression for $D$  in the form  
$
 \bar{J} \alpha^{-1}\mathring{D} + \bar{\dot{\by}}\cdot\mathring\bn
$
where $\bar{J}=\det\overline{\nabla\by}$.  In the special case when the position of the growing surface is fixed in the actual configuration $(D=0)$ the  Eulerian growth rate takes a simpler  form $\dot m/\alpha =-\bar{\varrho} (\bar{\dot{\by}}\cdot\mathring\bn)$.
 
Finally we remark that if the functions  $\bar\by (\bx)$ are not constrained directly on the growth surface, we can always use the freedom of choosing the Lagrangian coordinates  of the arriving material particles  to set $\bar\by (\bx)=\bx$. This will correspond to the choice of the \emph{reference} configuration coinciding at the moment of deposition with the instantaneous \emph{actual} configuration and will partially  specify  the reference metric. Such ``pseudo-linearization'' of the  deformation geometry  would be of course  impossible in many realistic cases,  for instance,  in  problems involving  growth against   rigid  obstacles.

\section{Solving the growth problem \label{Sec4}}

We now show that  if the function $\mathring\vartheta(\bx)$ is given, the  six controls  $(\bar\by(\bx),{\S}_a(\bx))$,  combined with the classical mechanical conditions on the growth surface, are sufficient to ensure a particular reference metric distribution $\mathring\G(\bx)$ in the grown body. 

Suppose, for simplicity,   that   the constitutive relation $ \S=\S( \nabla\by,\mathring\G)$ can be inverted, allowing one to express functionally the (time independent)  reference metric in terms of the (time dependent) current values of stress and strain, to obtain  $\mathring\G=\mathring\G(\nabla\by,\S)$.  For instance, this will be the case when  the elastic  energy is quadratic in strain, which in the case of isotropy implies that
\beq\label{Hencky}
e(\G,\mathring\G) = \frac{E}{2(1+\nu)}\left( \tr (\beps_e^2) + \frac{\nu}{1-2\nu}(\tr \beps_e)^2\right).
\eeq
Here the elastic strain $\beps_e$ can be chosen in different ways, for instance, in the   ``linear''  form 
$
\beps_e^{\text{lin}}=(\G-\mathring\G)/2
$
defining Kirchhoff-S.Venant  material \cite{GreenNaghdi,Efrati}, or  in the ``logarithmic''  form 
$
\beps_e^{\text{log}}=\left(\log(\G\mathring\G^{-1})\right)/2
$
 defining  Hencky material \cite{Bruhns,Auricchio,Anand}.  Note that in the case of   linear strain $\beps_e^{\text{lin}}$, a finite  energy is needed to squeeze a volume element  into a point, while this energy is infinite  when we use  the logarithmic strain $\beps_e^{\text{log}}$.  

The obtained  inversion $ \S=\S( \nabla\by,\mathring\G)$, however,   does not  solve our problem directly, because the projection of such  relation on the growth surface involves $\overline{\nabla\by}$ rather than $\nabla\bar\by$. Indeed, the (commutation) relation \eqref{com}
  indicates an essential coupling of the equilibrium problems \eqref{equires2} at different time instants. The functions $\overline{\dot{\by}}(\bx)$ will then depend  on  the whole deposition protocol,   making the relation between  $\mathring\G(\bx)$ and the controls $(\bar\by (\bx), \S_a (\bx))$ fundamentally \emph{nonlocal}. The simplest way to deal with such path dependence is to consider  an \emph{incremental} reformulation of the problem.

We start with the straightforward representations 
\beq\label{Sdot}
\begin{array}{ccc}
\displaystyle\nabla\by (\bx,t) = \overline{\nabla\by} (\bx) + \int_{\mathring\vartheta (\bx)}^t\nabla\dot\by(\bx,s)\,ds\\
\displaystyle\S(\bx,t) = \overline{\S}(\bx) + \int_{\mathring\vartheta (\bx)}^t\dot\S(\bx,s)\,ds. 
\end{array}
\eeq
In \eqref{Sdot} we can use the relation $\dot\S=\mathbfcal{A}\,\dot\nabla\by$  
 where $\mathbfcal{A}=\partial_{\nabla\by}^2 e(\G,\mathring\G)$  is the tangential elasticity tensor, which depends on the reference metric. We can also write  $\Div\S=\Div\bar\S + \int_{\mathring\vartheta(\bx)}^t\Div\dot\S\,ds - \dot\S\nabla\mathring\vartheta$, and since body forces can be   written as
\beq
\f(\bx,t) = \bar\f(\bx) + \int_{\mathring\vartheta(\bx)}^t\dot{\f}(\bx,s)\,ds
\eeq
the equilibrium condition $\Div\S+\f=\0$ transforms into  the incremental field equation for the stress increment, together with a boundary condition on incremental tractions \cite{Trincher,ZTprl}
\beq\label{dot3D}
\left\{
\begin{array}{lcc}
\Div\dot\S + \dot\f = \0 &&  \mathring\Body_t\\
\dot\S\mathring\bn  =  \mathring{D}\left(\Div\overline{\S} + \overline{\f}\right)  &&  \partial \mathring\Body_t. 
\end{array}
\right.
\eeq
The boundary condition \eqref{dot3D}$_2$ shows that if the new mass points are deposited on $\partial \mathring\Body_t$ in a state of mechanical unbalance ($\Div\overline{\S}+\overline\f\neq 0$), they are instantly re-equilibrated by activating  the incremental elastic displacements $\dot\by(\bx,t)$.

 The knowledge of   the functions $ \nabla\by (\bx,t)$ and $ \S (\bx,t)$ at each time step allows one to use the relation  $\mathring\G=\mathring\G(\overline{\nabla\by},\overline\S)$ to compute the reference metric of the newly adhered layers. This information is needed to update the  tangential elasticity tensor $\mathbfcal{A}$, which can be then used  to solve  the next incremental equilibrium problem.

\section{Linearization\label{Sec5}}
 
 It is instructive to compare  the nonlinear theory presented above with its linear counterpart \cite{ZTprl,ZTMaugin}.  Suppose that the displacement  $\bu(\bx,t) = \by(\bx,t)- \bx$ is small in the usual sense of linear elasticity. Then the analysis of the current metric $\G(\bx,t)$ can be replaced  by the study of the linearized strain $\beps(\bx,t)=\sym \nabla\bu(\bx,t)$, and instead of the reference metric $\mathring\G(\bx)$ one can consider  the linear reference strain $\mathring\beps(\bx)$. In this approximation, the elastic constitutive relation takes the simple form $\S= \bbC\beps_e$, where  $ \beps_e=\beps-\mathring\beps$ is the elastic strain. The Hookean elasticity tensor $\bbC$ will be  taken positive definite and space/time independent. 

While in the nonlinear theory the function $\bar{\by}(\bx)$  can be appreciably different from $\bx$, in the linear setting we must have  $\bar{\by}(\bx) \simeq \bx$. This implies that out of our six controls, three are automatically fixed and we have   only  the surface stress ${\S}_a(\bx)$ to work  with.  This is, of course, consistent with the fact that  in the linear theory only the three independent components of the incompatibility  of the reference strain $\mathring\bet=\Curl\Curl\mathring\beps$   affect the solution of the elasticity problem \cite{VanGoethem}. A fundamental  advantage of the linear formulation is that the relation between the incompatibility in  the newly accreted points  and the solution of the incremental problem can be made explicit  \cite{ZTprl}
\beq\label{lineta}
\mathring\bet = \overline\bet - \Curl\Curl\,(\sym(\overline{\dot\bu}\otimes\nabla\mathring\vartheta)),
\eeq
where the tensor $\overline\bet=-\Curl\Curl(\bbC^{-1}\overline{\S})$ can be viewed as a measure of  the incompatibility in the arriving material. If instead the reference incompatibility $\mathring\bet$ is known, the stress distribution at each moment of time $t$ can be found directly by solving the boundary value problem 
\beq\label{stressformlin}
\left\{
\begin{array}{lll}
\Div\S + \f = \0 & & \mathring\Body_t\\
\Curl\Curl (\bbC^{-1}\S) +\mathring\bet = \0 & &  \mathring\Body_t\\
\S\bn = \bs & & \partial\mathring\Body_t.
\end{array}
\right.
\eeq
Such a formulation is, of course, not possible in the nonlinear setting, where the knowledge of the full tensor $\mathring\G(\bx)$ is required to find the stress distribution  $\S(\bx,t)$ at each stage   of mass accretion.

In the incremental problem  the equilibrium equations \eqref{dot3D} remain the same with $\dot\S(\bx,t) = \bbC\nabla\dot\bu(\bx,t)$. An  important difference, though, is that the incremental displacement field is defined on a fixed configuration of the body,  which is controlled  at each moment of time by the function $\mathring\vartheta(\bx)$. Otherwise the  procedure of solving a one parametric family of the incremental problems remain the same as in the nonlinear case  with  prescription of  the tensor  $\overline\S_p(\bx) $ which defines the tractions $\bs(\bx)$  and fixes the three extra  components of surface stress   $\overline\S_a(\bx)$  controlling  the residual incompatibility $\mathring\bet(\bx)$. 

While the linearized theory preserves some  of the complexity  of the full nonlinear formulation, it under-represents   several important effects.   For instance, in the linear theory the reference and actual configurations are identified, which  makes it impossible to deal with problems involving confined growth in fixed domains.   Another problem is that in the linearized theory we  assume that the  elasticity moduli are fixed, while in the nonlinear theory,  the whole deposition history is  encoded both in $\mathring\G$, and in the tensor field $\mathbfcal{A}(\bx,t)$.   The implied  elastic inhomogeneity, accumulated during accretion, is particularly relevant for  biological materials where the hardening nonlinearity may be extremely strong  \cite{MacKintosh}.  Strong inhomogeneity of  elastic properties   can also become  the origin  of material instabilities \cite{FosdickEdelstein}, that  can, of course,  serve  by themselves  as potential targets in the design of surface growth protocols.

\section{Radial symmetry\label{Sec6}}

To  illustrate the above general  theory,   consider now in more detail the accretion process under the condition of radial symmetry (both in two and in three dimensions). 

Assume that the evolving reference configuration is a part of a ball $\mathring\Body_t= \left\{\bx\,|\,\psi_0\leq r \leq\psi_t \right\}$ in $\textbf{R}^n$ where $r=||\bx||$.  The function $\psi_t=\psi_0+\mathring{D}\,t$ will define the   position of the growth surface at  time $t$,  where $\psi_0$ is its initial position. For simplicity, the referential growth velocity $\mathring{D}$ will be taken as  constant. We can then write  $\mathring\vartheta(r) = (r-\psi_0)/\mathring{D}$, where $\mathring{D}=1/ \mathring{\vartheta}'$. 

Denote the unit vector pointing in radial direction by $\be=\bx/r$. Then the radially symmetric deformation can be written as $\by(\bx,t) = \chi(r,t)\be(\bx)$. The ensuing deformation gradient is
\beq\label{deff}
\nabla\by(\bx,t) = \lambda_r(r,t)\be(\bx)\otimes\be(\bx) + \lambda_{\theta}(r,t)\bbP(\bx)
\eeq
where $\lambda_r(r,t) =\partial_r  \chi(r,t)$ and $\lambda_{\theta}(r,t) =  \chi(r,t)/r$ are the stretches in radial and azimuthal directions,  and $\bbP=\I-{\be}\otimes{\be}$  is the projection tensor.  In view of radial symmetry, the most general form of the time-independent reference metric is
\beq\label{Gsph}
\mathring{\G}(\bx) = \gamma_r^2(r)\be(\bx)\otimes\be(\bx)  + \gamma_{\theta}^2(r)\bbP(\bx). 
\eeq
 For the case of a disk ($n=2$), the Ricci tensor of $\mathring\G$ reduces to  the Ricci scalar (twice the Gaussian curvature $K$)
 \cite{Kuhnel}  
\beq\label{Sd1}
S^d=2K= 2\frac{\gamma_{\theta}\gamma_r'+r\gamma_r'\gamma_{\theta}'-\gamma_r(2\gamma_{\theta}'+r\gamma_{\theta}''))}{r\gamma_r^3\gamma_{\theta}}. 
\eeq
The requirement $S^d=0$ produces then  a single differential constraint
 \beq\label{Sd0}
\gamma_r=k(r\gamma_{\theta})'
\eeq
 on the functions $(\gamma_r,\gamma_{\theta})$ where $k$ is an arbitrary constant. For a 3D sphere ($n=3$)  the Ricci tensor reduces to two independent components, 
\beq
\left\{
\begin{array}{lll}
R_1^s= 2\frac{\gamma_{\theta}\gamma_r'+r\gamma_r'\gamma_{\theta}'-\gamma_r(2\gamma_{\theta}'+r\gamma_{\theta}'')}{r\gamma_r\gamma_{\theta}}\\
R_2^s=\frac{\gamma_r^3+r\gamma_{\theta}\gamma_r'(r\gamma_{\theta})'-\gamma_r(\gamma_{\theta}^2+r^2\gamma_{\theta}'^2+r\gamma_{\theta}(4\gamma_{\theta}'+r\gamma_{\theta}''))}{\gamma_r^3}
\end{array}
\right. 
\eeq
Such Ricci tensor vanishes if and only if $\gamma_r=\gamma_{\theta}=\text{const}$, which corresponds to a homogeneous conformal dilatation of the reference sphere.

To determine  the unknown functions $\gamma_r(r)$ and $\gamma_{\theta}(r)$ we need to prescribe two auxiliary conditions on the growth surface.  In the case of radial symmetry   (for both $n$= 2, 3)  the vector function $\bar\by$ reduces   to a  scalar function $\bar{\chi}(r)$ and, in view of  the representation $\bar{\S}=\bar{s}_r(r)\be\otimes\be + \bar{s}_{\theta}(r)\bbP$, the active   stress $\S_a$ reduces  to its hoop  component $\bar{s}_{\theta}(r)$.  The additional  conditions prescribing the growth protocol,  may then  take the form of restrictions imposed on the functions $\bar{\chi}(r)$ and $\bar{s}_{\theta}(r)$ with   the other component of the deposition stress $\bar{s}_r(r)$  fixed passively. Note that  in the problem of inward accretion against a fixed wall we have  $\mathring\bn=-\be$, so that the mass flux $\dot m =\bar{\varrho}\alpha\bar{\dot{\chi}}$;  for outward accretion, $\mathring\bn=\be$ and  $\dot m =-\bar{\varrho}\alpha\bar{\dot{\chi}}$. 
 
 To formulate the incremental problem  when  the functions $\bar{\chi}(r)$ and $\bar{s}_{\theta}(r)$ are   known, we first need to specialize \eqref{Sdot} for the case of spherical symmetry 
\beq\label{incrementalall}
\begin{array}{cc}
\displaystyle\lambda_{r/\theta}(r,t) = \bar{\lambda}_{r/\theta}(r) + \int_{\mathring\vartheta(r)}^t\dot{\lambda}_{r/\theta}(r,s)\,ds\\
\displaystyle s_{r/\theta}(r,t) = \overline{s}_{r/\theta}(r) + \int_{\mathring\vartheta(r)}^t\dot{s}_{r/\theta}(r,s)\,ds
\end{array}
\eeq
where  the deposition stretches  can be written as 
\beq\label{ylambdas}
\left\{
\begin{array}{lll}
\bar\lambda_r(r) =\bar{\chi}'(r)  - \mathring{\vartheta}'(r)\,\overline{\dot{\chi}}(r)\\
\bar\lambda_{\theta}(r)=\bar{\chi}(r)/r.
\end{array}
\right.
\eeq
To compute the rates  $\dot\lambda_r=\partial\dot{\chi}$, $\dot\lambda_{\theta} = \dot{\chi}/r$ we need to  introduce the incremental moduli
\beq\label{dotsnew}
\left\{
\begin{array}{llll}
\displaystyle\dot s_r ={\mathcal{A}}_{rr}\dot{\lambda}_r +{\mathcal{A}}_{r\theta}\dot{\lambda}_{\theta}\\
\displaystyle\dot s_{\theta} = {\mathcal{A}}_{\theta r}\dot{\lambda}_r + {\mathcal{A}}_{\theta\theta}\dot{\lambda}_{\theta}. 
\end{array}
\right.
\eeq
These moduli can  be  expressed in terms of the current values of stretches $ \displaystyle\lambda_{r/\theta}(r,t)$ and stresses $\displaystyle s_{r/\theta}(r,t)$ which, as we have mentioned before,  depend on the whole accretion history.

Since $\mathring\G(\bx)$ is  time-independent, we can  time differentiate  the inverse constitutive equation $\mathring\G(\nabla\by,\S)$ to  obtain the relation $ \partial_{\nabla\by}\mathring\G[\nabla\dot\by] + \partial_{\S}\mathring\G[\dot\S]=0$. Then $\dot\S=\tilde{\mathbfcal{A}}\nabla\dot\by$, where  $\tilde{\mathbfcal{A}}=-(\partial_{\nabla\by}\mathring\G)^{-1}(\partial_{\S}\mathring\G)$. In the  case of a sphere, we can first use the elastic model \eqref{Hencky}  with  linear strains $\beps_e^{\text{lin}}$  to invert the elastic constitutive relation    
\beq\label{Linearsph} 
\left\{
\begin{array}{lll}
\displaystyle\gamma_r=\sqrt{\lambda_r^2 - 2 s_r/(E\lambda_r) + 4 s_{\theta}\nu/(E\lambda_{\theta}))}\\
\displaystyle\gamma_{\theta}=\sqrt{\lambda_{\theta}^2+2 s_r\nu/(E\lambda_r) + 2 s_{\theta}(\nu-1)/(E\lambda_{\theta})}. 
\end{array}
\right.
\eeq
If, instead, we use the  logarithmic strain $\beps_e^{\text{log}}$, we obtain
\beq\label{Henckysph} 
\left\{
\begin{array}{lll}
\displaystyle\gamma_r=\lambda_r\exp\left(\left(2\nu s_{\theta}\lambda_{\theta} - s_r\lambda_r \right)/E\right)\\
\displaystyle\gamma_{\theta}=\lambda_{\theta}\exp\left(\left(s_{\theta}\lambda_{\theta}(\nu-1) + \nu s_r\lambda_r\right)/E\right). 
\end{array}
\right.
\eeq
Since the inelastic strains are time-independent, by time differentiation of \eqref{Linearsph} and \eqref{Henckysph}  we obtain the incremental constitutive equations. In the case when we use the  linear strain measure $\beps_e^{\text{lin}}$, the tangential elasticity tensor $\tilde{\mathbfcal{A}}$ has components
\beq\label{BBB}
\left\{
\begin{array}{llll}
\displaystyle\tilde{\mathcal{A}}_{rr} =E\lambda_r^2(1-\nu)/((1+\nu)(1-2\nu)) + s_r/\lambda_r \\
\displaystyle\tilde{\mathcal{A}}_{r\theta} = \tilde{\mathcal{A}}_{\theta r} = E\nu\lambda_r\lambda_{\theta}/((1+\nu)(1-2\nu))\\
\displaystyle\tilde{\mathcal{A}}_{\theta\theta} =E\lambda_{\theta}^2(1-\nu)/((1+\nu)(1-2\nu)) + s_{\theta}/\lambda_{\theta} 
\end{array}
\right.
\eeq
while if we use the logarithmic strain $\beps_e^{\text{log}}$,
\beq\label{AAA}
\left\{
\begin{array}{llll}
\displaystyle\hat{\mathcal{A}}_{rr} = E(1-\nu)/(\lambda_r^2(1+\nu)(1-2\nu)) - s_r/\lambda_r \\
\displaystyle\hat{\mathcal{A}}_{r\theta} = \hat{\mathcal{A}}_{\theta r} = E\nu/(\lambda_r\lambda_{\theta}(1+\nu)(1-2\nu)) \\
\displaystyle\hat{\mathcal{A}}_{\theta\theta} =   E(1-\nu)/(\lambda_{\theta}^2(1+\nu)(1-2\nu)) - s_{\theta}/\lambda_{\theta}.
\end{array}
\right.
\eeq
We now have all we need to formulate the  sequence of incremental equilibrium problems. Specialization of the equilibrium equations \eqref{dot3D}$_1$  to the case of radial  symmetry  gives
\beq\label{eqdotspher} 
\frac{\partial\dot{s}_r}{\partial r}+ \frac{n-1}{r}(\dot{s}_r - \dot{s}_{\theta}) = 0.
\eeq
The boundary condition \eqref{dot3D}$_2$  on the growth surface takes the form 
\beq\label{bndnongrow}
 \dot{s}_r = \mathring{D}\left(\frac{d\overline{s}_r}{dr}+ \frac{n-1}{r}(\overline{s}_r - \overline{s}_{\theta}) \right). 
\eeq
Note that often it may be more  convenient to specify the deposition protocol in  Eulerian  rather than Lagrangian coordinates. In this case we can still formulate the incremental problem in terms of Piola-Kirchhoff components of stress, however,  the boundary conditions \eqref{bndnongrow} would have to be modified.  Thus, if  we can control at deposition the components of  Cauchy rather than Piola-Kirchhoff  stress, the incremental boundary condition describing the radially symmetric  growth  will take the form 
\beq
\label{Trdisk}
\dot{s}_r - \frac{1}{r}\overline{\s}_{\theta}\overline{\dot\chi}=  D\left(\overline{\s}'_r\bar\lambda_{\theta} + \frac{1}{r}(\overline\s_r-\overline\s_{\theta})\right)
\eeq
in the case  of  a  disk ($n=2$), and 
\beq
\label{Trsph}
\dot{s}_r - \frac{2}{r}\overline{\s}_{\theta}\bar\lambda_{\theta}\overline{\dot\chi}= D\left(\overline{\s}'_r\bar\lambda_{\theta} + \frac{2}{r}(\overline\s_r-\overline\s_{\theta})\right)\bar\lambda_{\theta} 
\eeq
in the case  of  a  sphere ($n=3$).   Here we   used the Eulerian velocity of the growing surface   $D=\bar{\chi}'\mathring{D}$, and we   denoted $\bar\s_r'(y)=d\s_r(y)/dy.$  In the case of a  disk ($n=2$) the relation between Cauchy and Piola-Kirchhoff stresses reads $s_r=\s_r\lambda_{\theta}$ and $s_{\theta}=\s_{\theta}\lambda_r$, while in the case of a sphere ($n=3$),   $s_r=\s_r\lambda_{\theta}^2$ and $s_{\theta}=\s_{\theta}\lambda_r\lambda_{\theta}$.  Note that  the boundary conditions \eqref{Trsph}, \eqref{Trdisk}  couple the controlled  stress   $\overline\s_r, \overline{\s}_{\theta}$  and  the controlled shape  $\bar{\chi}(r)$, which is a purely nonlinear effect. 

\section{Case studies\label{Sec7}}

To illustrate different aspects of the developed general theory,  we now elaborate few explicit  case studies, which are all relevant for applications. Our main  goal here  is to highlight the effects of physical and geometrical nonlinearities, and to illustrate path dependence of the surface growth process.  Our examples show that the deposition protocol is effectively \emph{remembered} by the grown body through the accumulated incompatibility, the inhomogeneity of the incremental  moduli and the  shape of the  body  liberated from the constraints.  In all   examples  we assume that the function $\mathring\vartheta(\bx)$ is prescribed, leaving aside the problem of the feedback received by the growth mechanism from the current state of stress.  

\subsection{Winding of 2D disks} 

\begin{figure}[thb] 
\includegraphics[width=8.5cm]{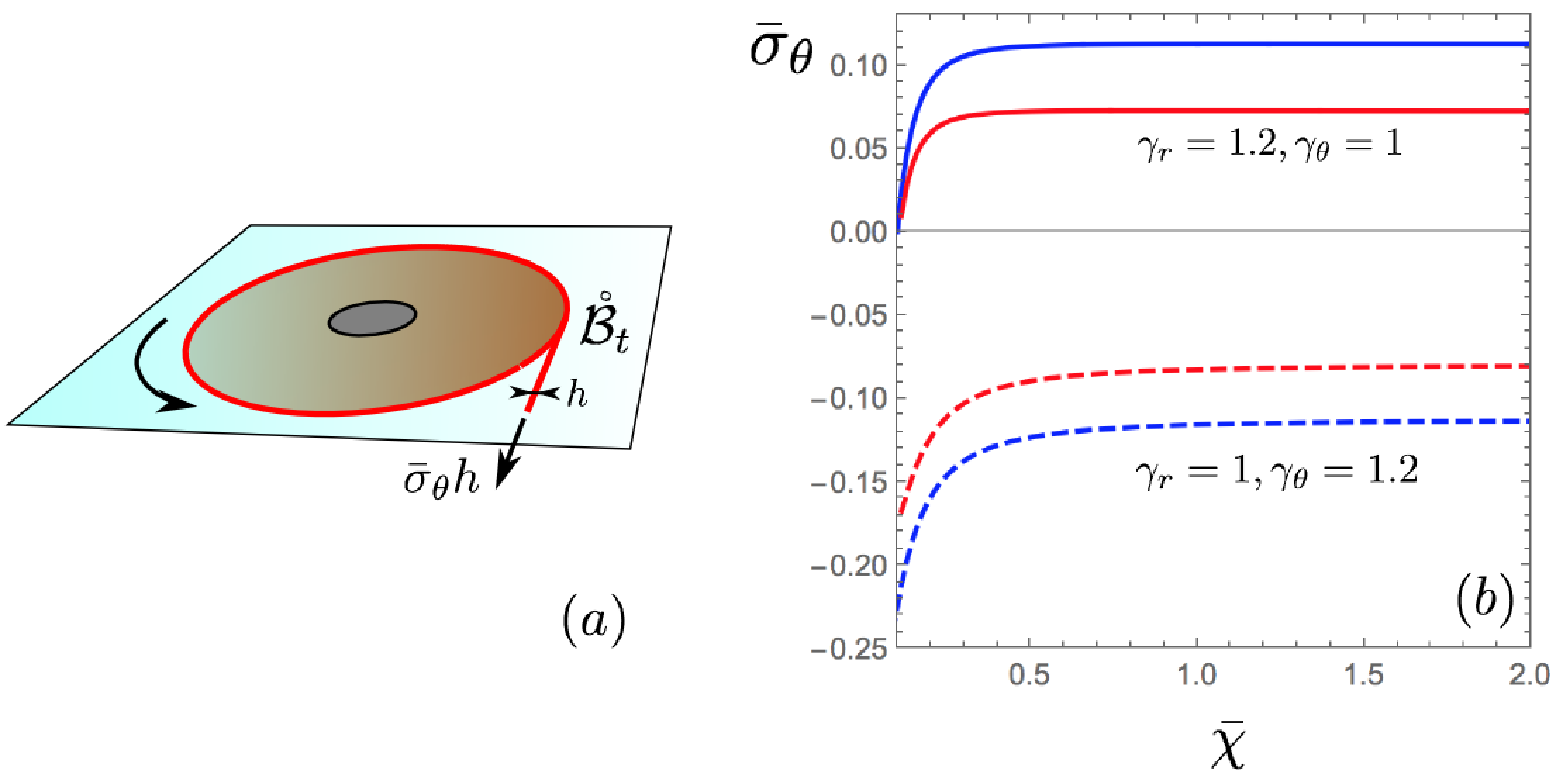}
\caption{\label{protocol} $(a)$ Schematic presentation of the 2D winding growth set up with controlled traction force (b) Winding protocols required to embed the diagonal reference metrics with: $\gamma_r=1.2$, $\gamma_{\theta}=1$ (solid) and  $\gamma_r=1$, $\gamma_{\theta}=1.2$ (dashed). Red - nonlinear Hencky model with $E=1$ and $\nu=0.3$:  blue  -  its linearization. In all plots $a=0.1$ and $r_e=2$.}
\end{figure}

Consider a process of  winding of an infinitesimally thin tape on a rigid mandrel. Suppose that the growth process takes place in 2D so that the emerging hollow disk is constrained to remain in plane.  Assume that the winding   is accomplished  by  pulling the adhering tape with a controlled tangential force. If the adhering layers have thickness $h$ and the  tangential force is  $F$,  we assume that in the double limit $h \to 0$, $F \to 0 $ the hoop stress $\bar\s_{\theta}=F/h$ remains finite, see Fig.\ref{protocol}a. The growing disk $\mathcal{D}=\{\bx\,|\,a\leq||\br||\leq\psi\}$ (with $\psi \leq r_e$) can be parameterized by the referential radius $\psi$ of the external circle where the tape deposition takes place.

We begin  with the simplest inverse problem: given that the acquired reference metric $\mathring\G=\text{diag}(\gamma_r,\gamma_{\theta})$ is homogeneous, find the corresponding deposition protocol.  Somewhat counter-intuitively,   this metric is nontrivial,  being 
 in general locally compatible and globally incompatible. Indeed,  in view of \eqref{Sd0} the Gaussian curvature $K$ is equal to zero everywhere except  for the origin, where the polar coordinate system is singular and the curvature has a delta function type singularity as long as  $k=\gamma_r/\gamma_{\theta}\neq 1$. Physically such target configuration corresponds to a disk with   removed   ($k>1$) or inserted  ($k<1$)  wedge (with opening angle $\D\theta=2\pi(k^{-1}-1)$) \cite{RazPRE,RazARMA}.

Observe next that the prescription of a reference metric implicitly constrains the choice of the Lagrangian coordinates for the reference configuration of the growing body.  If we  assume, in addition,  that the normal tractions on the growth surface are absent  ($s_r(\psi,\psi) = 0$), we can solve the sequence of incremental problems and compute \emph{both}, the  current  hoop stress $\bar s_{\theta}(\psi)=s_{\theta}(\psi,\psi)$ and the current radius of the disk $\bar\chi(\psi)$.  With this information at hand,  we can express the Cauchy hoop stress $\s_{\theta}=s_{\theta}/\lambda_r$  in terms of the current position of the grow surface, to obtain the function $\s_{\theta}(\chi)$ controlling the winding process.  Our computational results are illustrated in Fig.\ref{protocol}b  for the case of 2D Hencky material with logarithmic strain $\beps_e^{\text{log}}$. 

Note that   to obtain discs with positive embedded Gaussian curvature  ($\gamma_r>\gamma_{\theta}$) the controlled traction force must be positive, whereas to embed  negative curvature ($\gamma_r<\gamma_{\theta}$),  it must be negative.  This is consistent with the fact that  disks with positive curvature  are obtained by removing a wedge: indeed,  by wrapping through pulling we deposit less mass per length  than in  the case of zero force wrapping.  Instead, pushing produces  an opposite result and leads to disks with negative curvature. 

To understand these numerical results more deeply, we can  compare them  with the  analytically transparent linearized theory. To this end we set  $\gamma_{r/\theta}=1+\mathring\e_{r/\theta}$ and  assume that the inelastic strains $\mathring\e_{r/\theta}$ are small. The linearization of the scalar Ricci curvature \eqref{Sd1} gives  the  compatibility condition in the bulk, 
\beq \label{compat}
{\mathring\e_{\theta}}''+\frac{2{\mathring\e}'_{\theta} - {\mathring\e}'_{r}}{r} = 0
\eeq
which can be also rewritten as  $(r\mathring\ph)'=0$, where $\mathring\ph={\mathring\e_{\theta}}'+({\mathring\e}_{\theta} - {\mathring\e}_{r})/r$. Therefore $\mathring\ph=c/r$ where  $c$ an arbitrary constant and if $c\neq 0$ there will be a curvature singularity at the origin (signifying the presence of an inserted or removed wedge). 

We reiterate that  even though  the inelastic strains $\mathring\e_{r/\theta}$ are homogeneous,   the residual stresses in this case will be different from zero.  Indeed, if  $u$ is a radial displacement,  the total strains $\e_r=u'$ and $\e_{\theta}=u/r$ can be  additively  decomposed as $\e_{r/\theta}=e_{r/\theta}+\mathring\e_{r/\theta}$, where  the elastic strains $e_{r/\theta}$ are constitutively related to  stresses  $\e_{r/\theta}=(\s_{r/\theta} - \nu\s_{\theta/r})/E$, where  $\nu$ is the Poisson ratio and $E$ is the  Young modulus. Since in equilibrium   $\s_r'+(\s_r-\s_{\theta})/r=0$, the residual stresses in a traction-free disk can be determined by solving  the equation $(r^3\s_r')' = - Er^2 \mathring\ph$ with the right hand side $\mathring\ph=({\mathring\e}_{\theta} - {\mathring\e}_{r})/r$. If  ${\mathring\e}_{\theta} \neq  {\mathring\e}_{r}$ the solution will be obviously nonzero even  for a hollow disk  with  zero  boundary conditions $\s_r(a) = \s_r(r_e) = 0$. 

We can now use the linearized theory to  address the peculiar behavior of the winding tension $\s_{\theta}(\psi)$ in the nonlinear theory,  see Fig.\ref{protocol}a.  Solving the linear equilibrium problem with boundary conditions $u(a,\psi)=0$ and $\s_r(\psi,\psi)=0$, we obtain $\s_{\theta}(\psi) = E((\mathring{\e}_r-\mathring{\e}_{\theta})(\psi^2-2a^2(1-2\nu)\log(a/\psi))-a^2(\mathring{\e}_r+\mathring{\e}_{\theta}(3-4\nu)))/(2(a^2(1-2\nu)+\psi^2)(1-\nu^2))$. In particular, at  $\psi\rightarrow\infty$  we have 
\beq\label{sigmalimit}
\s_{\theta}^{\infty} = \frac{E(\mathring{\e}_r-\mathring{\e}_{\theta})}{2(1-\nu^2)}
\eeq 
whereas in the limit $\psi\rightarrow a$ we obtain
\beq\label{sigmamandrel}
\s_{\theta}^{a} = - \frac{E\mathring{\e}_{\theta}}{1-\nu^2}. 
\eeq 
These analytical results confirm the general trends observed in the numerical solution of the  nonlinear problem and show to what extent  the linearized theory   overestimates the necessary winding tension, see Fig.\ref{protocol}. 
  
 \begin{figure}[thb] 
\includegraphics[width=6.5cm]{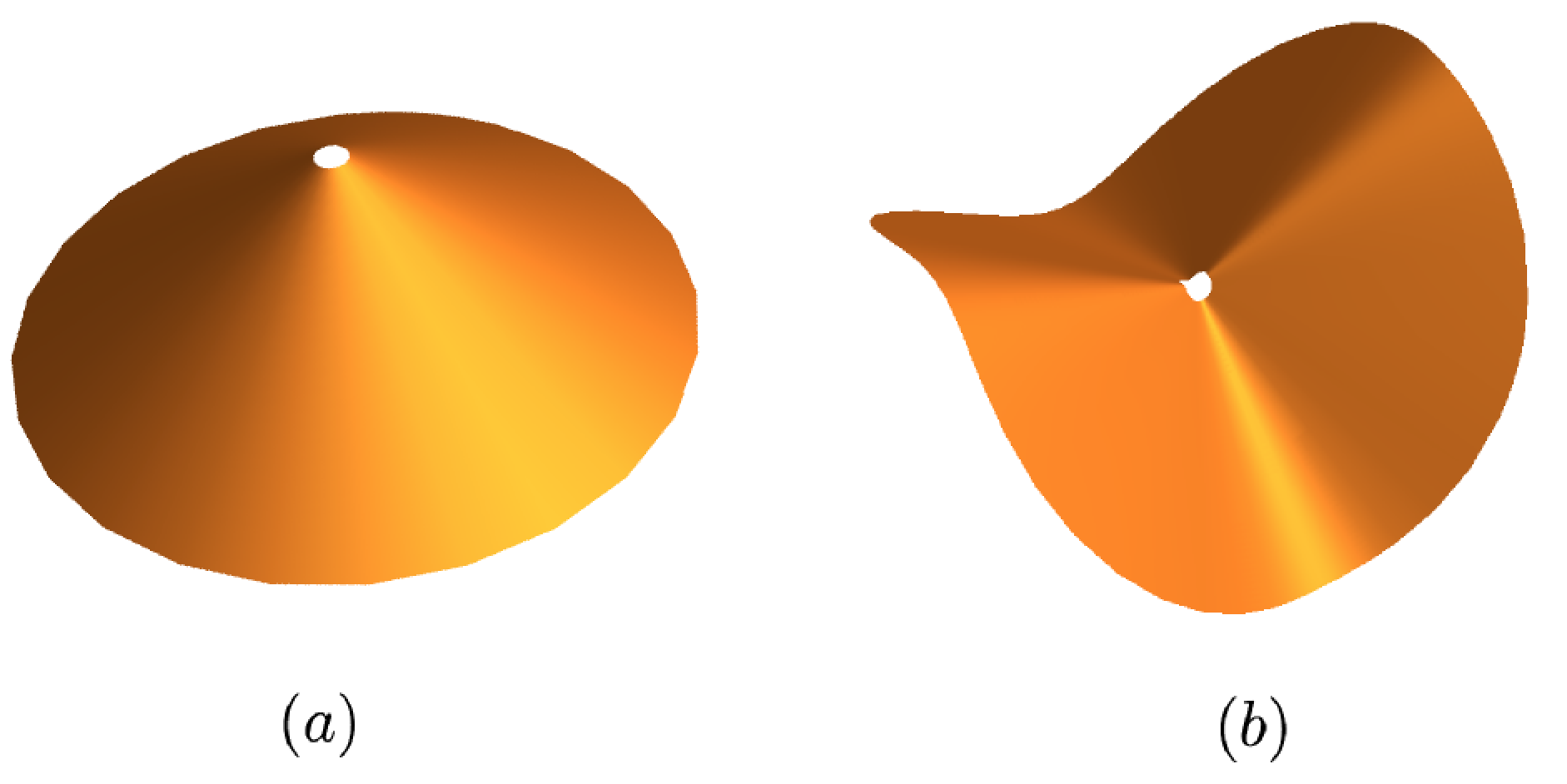}
\caption{\label{K1} Stress-free 3D configurations of prestressed 2D disks: $(a)$  exact result for $\gamma_r=1.2$ and $\gamma_{\theta}=1$.  $(b)$ approximation with $m=2$, $A=0.55$ for $\gamma_r=1$ and $\gamma_{\theta}=1.2$.}
\end{figure}

 \subsection{Isometric embedding of  2D disks into 3D}

To illustrate the growth-induced  incompatibility produced by winding,  we now allow  the grown disk to detach from its (imaginary) 2D substrate and take a relaxed shape  in 3D.  Such an  isometric embedding  $\bpsi:\bbR^2\mapsto\bbR^3$ leads to the  full relaxation of elastic energy  and must therefore satisfy the system of equations   $\mathring\G=\nabla\bpsi\trsp\nabla\bpsi$,  where $(\gamma_r,\gamma_{\theta})$ is the homogeneous diagonal reference metric acquired in the process of controlled growth discussed above. We remark that such  relaxed shapes may serve by themselves  as the target of  controlled surface manufacturing \cite{Sharon, Danescu}. 

Note first that the relaxed  shape  of the disk in 3D  depends crucially on the sign of its Gaussian curvature  which is proportional to the effective wedge opening $\D\theta=2\pi(\gamma_{\theta}/\gamma_r - 1)$.  As we show below,  for $\gamma_r>\gamma_{\theta}$ the isometric embedding takes the form  a circular cone, whereas for $\gamma_{\theta}>\gamma_r$ it  is known as an {\it excess} cone  (or {\it anti }cone) \cite{BenAmarConicalDefects,Warner}. 

We begin with the trivial case  $\gamma_r=\gamma_{\theta}$, when the isometric embedding is confined to  2D and $\bpsi(\bx)=g(r)\be(\bx):\bbR^2\mapsto\bbR^2$. Here  $\be$ is the unit vector in the radial direction, and $g(r) = r \gamma_{\theta} = r \gamma_r$. Clearly, in this case, the relaxed configuration of the disk  is just  another flat  disk. 

If   $\gamma_r>\gamma_{\theta}$,  the Gaussian curvature is positive and, even if the functions  $\gamma_r$ and $\gamma_{\theta}$ are $r$ dependent,  we can write  a universal embedding   $\bpsi(\bx)=\rho(r)\be(\bx) + \zeta(r)\bk$ where the unit vector  $\bk$ is perpendicular to the reference plane.  The unknown functions $\rho(r)$ and $ \zeta(r)$ must satisfy  $\gamma_r^2={\rho'^2+\zeta'^2}$ and $\gamma_{\theta}=\rho/r$, so that for homogeneous metric $\rho=r \gamma_{\theta}$ and $\zeta=(\gamma_r^2 - \gamma_{\theta})^2)^{1/2}$; the latter expression shows that $\zeta$ is real only if $\gamma_r>\gamma_{\theta}$. The corresponding relaxed shape is shown   in Fig. \ref{K1}a.

\begin{figure}[thb] 
\includegraphics[width=4cm]{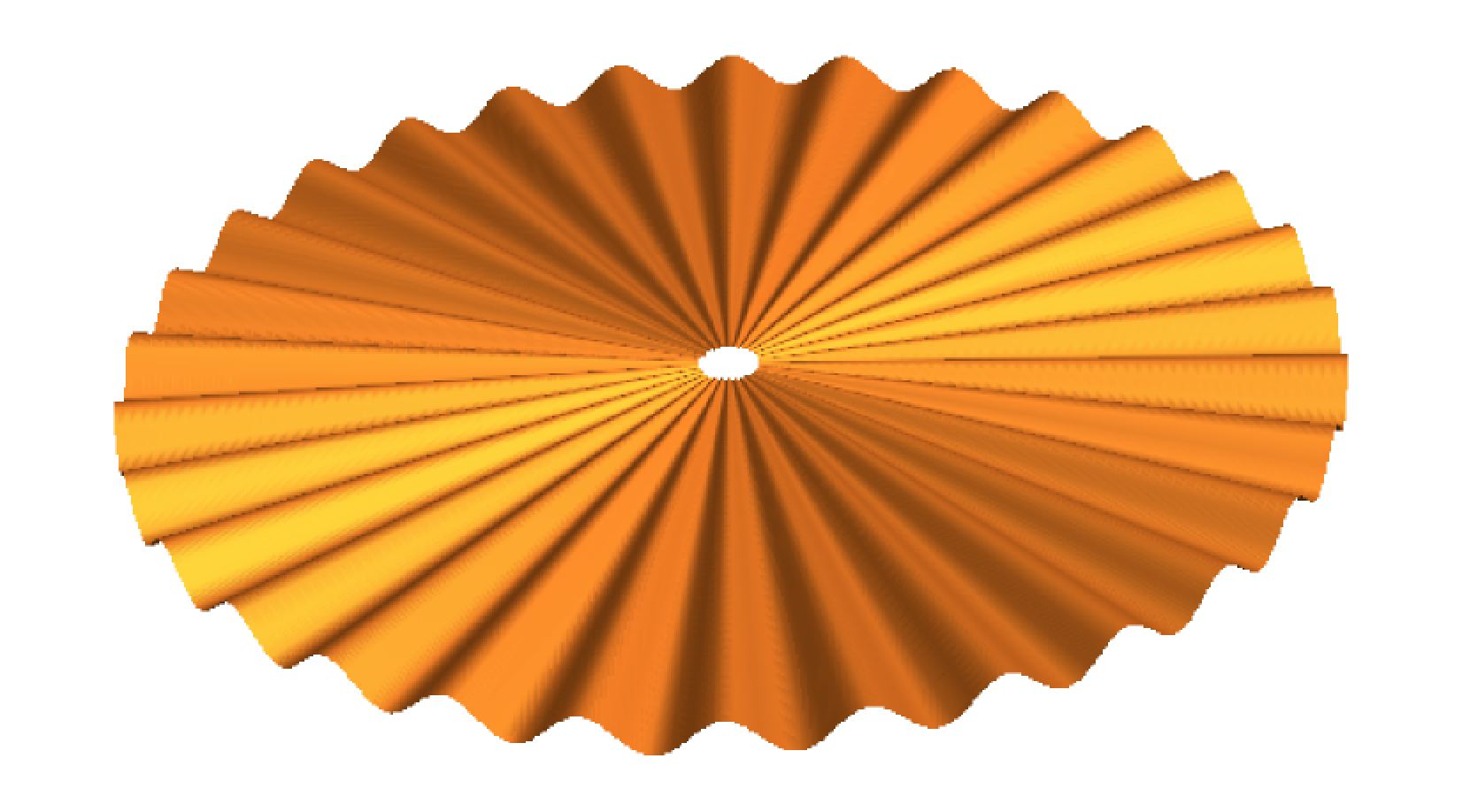}
\caption{\label{K2} Stress-free 3D configuration of a prestressed 2D disk with $\gamma_r=1$ and $\gamma_{\theta}=1.2$:  approximate solution with $m = 30,A = 0.032$.}
\end{figure}

When $\gamma_r<\gamma_{\theta}$ the embedded curvature is negative and  the reconstruction of the 3D relaxed surface is not straightforward \cite{BenAmarConicalDefects,Sharon,HanHong}. An approximate construction can be built  based on the auxiliary  surface   $\bpsi(\bx)=r\rho(\theta)\be(\bx) + r\zeta(\theta)\bk$ (see \cite{Warner}), whose induced metric is  also diagonal $(\tilde\gamma_r^2(\theta),\tilde\gamma_{\theta}^2(\theta))$  with 
\beq
\left\{
\begin{array}{lll}
\tilde\gamma_r^2(\theta) = \rho^2(\theta) + \zeta^2(\theta)\\
\tilde\gamma_{\theta}^2(\theta) = \tilde\gamma_r^2(\theta) (1+\zeta'^2(\theta)/(\tilde\gamma_r^2(\theta)-\zeta^2(\theta))) - \zeta^2(\theta). 
\end{array}
\right.
\eeq
Generically, this  metric  is not  homogeneous and  is therefore, strictly speaking,   incompatible with the homogeneous  reference metric generated by the winding process. However, we can demand that  one of its components is homogeneous  $\tilde\gamma_r(\theta)=\gamma_r$ and the other one meets the original anzats \emph{in average} so that
$
\frac{1}{2\pi}\int_0^{2\pi}\tilde\gamma_{\theta}(\theta)\,d\theta = \gamma_{\theta}.
$
To meet this constraint, whose  physical  meaning  is that the image of a constrained  reference circle has a prescribed  perimeter in the relaxed state \cite{Warner},  we can, for instance, assume that $\zeta(\theta)=A\,\text{sin}(m\theta)$, where $m>1$ is an integer. 

The integral condition provides a link between  $A$ and $m$ and  
for the case of homogeneous  $\gamma_r,\gamma_{\theta}$, the amplitude of the oscillation decreases with increasing wave number $m$, see Fig.  \ref{K2}. The resulting smooth  embedding is, of course,  only an approximation of the relaxed surface which  can be expected to be rather rough \cite{Nash}. However, lower order modes, like the one shown in   Fig. \ref{K1}b, will  be energetically favored  if, for instance,  small bending stiffness   is  taken into consideration.

\subsection{Outward growth under controlled pressure} 

Our  next example concerns  outward accretion of a hollow  sphere under controlled pressure, which is physically  relevant in the context of pressurized freezing \cite{Fletcher}.  The goal of this example  is to illustrate the  possibility of extreme  path dependence of the accretion process. 

We assume that  the reference configuration is   a hollow sphere  $\mathring\Body_t= \left\{\bx\,|\,A\leq||\bx||\leq\psi_t\leq B\right\}$, where $A$ and $B$ are the initial and final radii.  The current configuration is $\Body_t=\left\{\by\,|\,a\leq||\by||\leq \bar\chi(\psi_t)\right\}$,  which shows that  the internal, non-growing surface is  forced to remain on a rigid bead of radius $a$. 
 
\begin{figure}[thb] 
\includegraphics[width=\columnwidth]{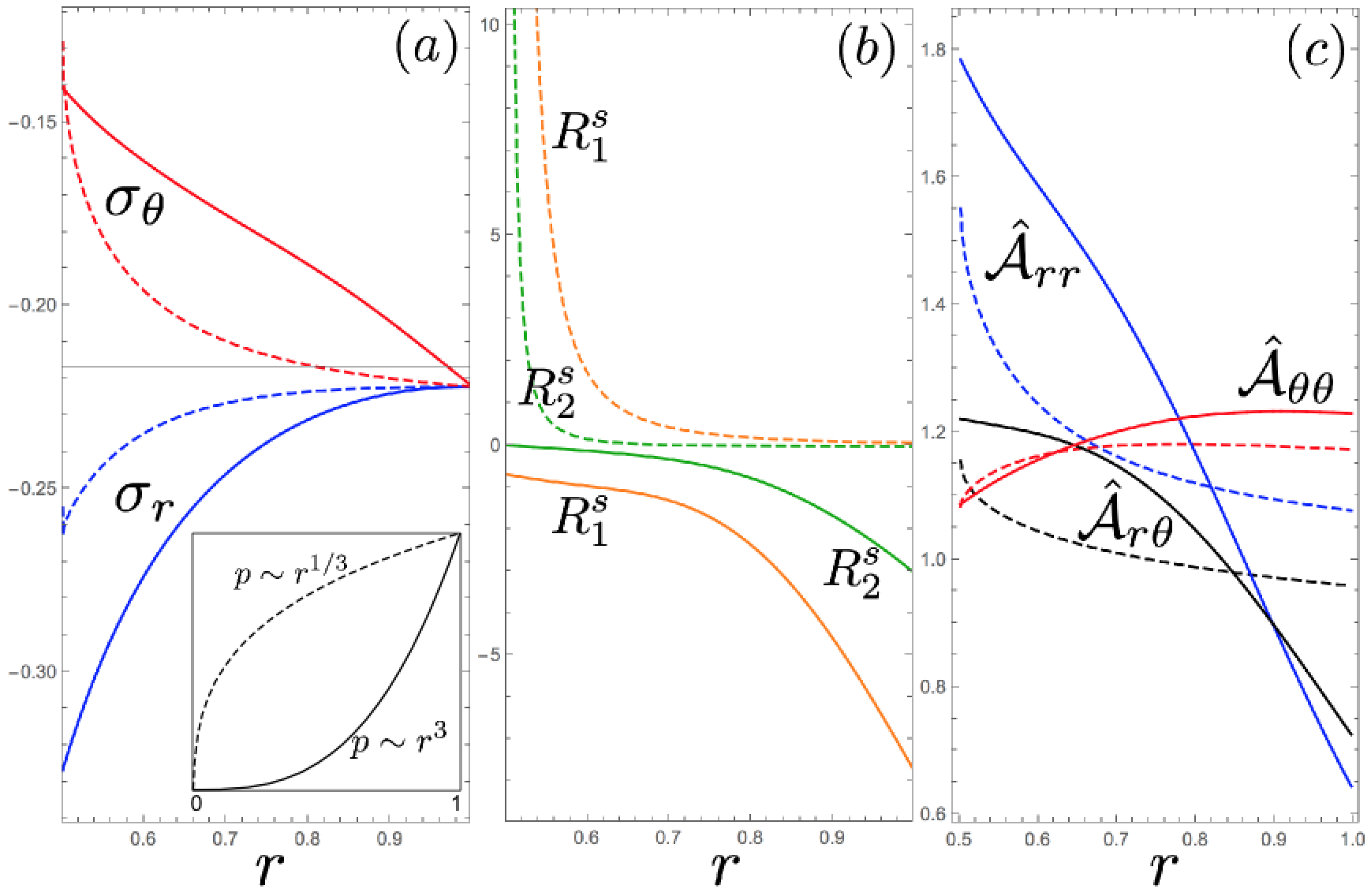}
\caption{\label{protocols} Effect of different pressure  protocols on the outcome of the outward accretion with controlled pressure. In all figures, solid curves corresponds to $m=3$ and dashed curves to $m=1/3$. $(a)$ blue -radial,  red - azimuthal components of stress  at the end of accretion process for the  two pressure protocols showed in the inset; $(b)$ the final distribution of the two components of Ricci curvature; $(c)$ the final  distributions of the elastic moduli; the color code is the same as in Fig.\ref{Misbah_plots}. The parameters: $\beta=0.2$Pa, $\nu=1/3$, $A=a=0.5$, $B=1$ .}
\end{figure}
 
We further assume that the growth protocol is characterized by the following  conditions on the accretion surface 
\beq\label{extrapress}
\left\{
\begin{array}{lll}
\bar\s_r(\psi_t) = \bar\s_{\theta}(\psi_t) = -p(\bar\chi(\psi_t)), \\
\bar{\chi}(\psi_t) = \psi_t.
\end{array}
\right.
\eeq
The second condition \eqref{extrapress}  is just the simplest assumption   that the reference configuration for the arriving material particles  coincides with the actual configuration. Instead, on the non-growing part of the boundary we impose the hard constraint $\chi(A,t)=A$. 

For determinacy, we  assume that  
\beq\label{protocolpress}
\displaystyle p(z) = \beta \left(\frac{z-\bar\chi(A)}{\bar\chi(B) - \bar\chi(A)}\right)^m
\eeq
where  $\beta$ and $m$ are prescribed constants. All members of this ``family'' begin  ($\psi_0\equiv A$) and end ($\psi_T\equiv B$) at the same  level  of pressure while   exhibiting  super or sub-linear growth  depending on  whether $m>1$ or $m<1$. 

Our numerical simulations for the Hencky material (with strain $\beps_e^{\text{log}}$) show appreciable  protocol (history) sensitivity, which can be seen in  the final stress distribution, see Fig. \ref{protocols}(a) and in the final  inhomogeneity of tangential elastic moduli, see Fig. \ref{protocols}(c). The most striking effect concerns the  final  distribution of incompatibility. As we see in   Fig. \ref{protocols}(b),   a  seemingly insignificant change in the control of pressure  during deposition,  can lead to either divergence of the two components of curvature at the inner boundary of the body (for $m<1$) or their  convergence to zero ( for $m>1$). In other words  our computations show that  the embedded curvature  lacks  continuity with respect to the parameter $m$. The origin of this dramatic effect  can be explained already   in the framework of a more transparent linearized theory. 

Note first that in 3D linearized theory  the metric is compatible if   two conditions are satisfied simultaneously: $(r\mathring\ph)'=0$ and $(r^3\mathring\ph)'=0$ where again $\mathring\ph=\mathring\e_{\theta}'+(\mathring\e_{\theta} - \mathring\e_r)/r$. The solution of these equations is $\mathring\ph=0$ and since the equilibrium problem for the stress distribution in the unloaded 3D body is
\beq \label{3D}
\left\{
\begin{array}{lll}
(r^4\s_r')' = - 2Er^3 \mathring\ph/(1-\nu)\\
\s_r(r_i) = \s_r(r_e) = 0, 
\end{array}
\right.
\eeq
 hollow spheres behave quite differently   than  hollow disks, where  there can  be residual stresses even if the regular part of curvature vanishes. 

The 3D linearized problem of surface accretion under a general  pressure distribution $p(z)$  can be solved explicitly. We now assume that $a=A$.  If we again replace  time   by the radius $\psi$ of the growing surface,   the solution to the problem \eqref{eqdotspher} with incremental boundary conditions $\dot{u}(a,\psi)=0$ and $\dot\s_r(\psi,\psi)=-p'(\psi)$ is  $\dot{u}(r,\psi)=c_1(\psi)r+c_2(\psi)/r^2$ with $c_1(\psi)=(1+\nu)(2\nu-1)\psi^3 p'(\psi)/(E(2(1-2\nu)a^3 + (1+\nu)\psi^3))$ and $c_2(\psi)=-a^3 c_1(\psi)$. The resulting stress distribution takes the form
\beq\label{stressspherelinear}
\left\{
\begin{array}{lll}
\s_r(r,\psi) = -p(r) + \int_r^{\psi}\frac{s^3(r^3(1+\nu)+2(1-2\nu)a^3)p'(s)}{r^3(s^3(1+\nu)+2(1-2\nu)a^3)}\,ds\\
\s_{\theta}(r,\psi) = -p(r) + \int_r^{\psi}\frac{s^3(r^3(1+\nu) - (1-2\nu)a^3)p'(s)}{r^3(s^3(1+\nu)+2(1-2\nu)a^3)}\,ds. 
\end{array}
\right.
\eeq
In particular, these expressions  show that the case  $a=0$ is trivial in the sense that  the stress in the growing sphere is necessarily hydrostatic.
 
The distribution of residual stresses in a hollow sphere of internal radius $r_i$ and external radius $r_e$ is given by \eqref{3D}. 
The expression for  $\mathring\ph(r)$ can be found by  solving the linearized problem of surface accretion   with  $r_i=a$:
\beq\label{curvsphere}
\mathring\ph(r) = \frac{3(1-\nu)(1-2\nu)a^3}{E r^3(1+\nu) + 2(1-2\nu)a^3}p'(r). 
\eeq
This expression  confirms  that  no incompatibility can result from  a growth  process unless the sphere is hollow ($a\neq 0$) and the applied pressure varies along the deposition process. If we substitute \eqref{protocolpress} into \eqref{curvsphere} and expand the result in small $\d=(r-a)/a$, we  obtain $\mathring\ph(\d) \sim \d^{m-1}$ which  shows that the curvature either diverges or   tends to zero in proximity of the inner radius depending on whether $m<1$ or $m>1$. This is exactly what we found numerically  in the case of nonlinear Hencky material (with strain $\beps_e^{\text{log}}$), see  Fig.\ref{protocols}b.

\begin{figure}[thb] 
\includegraphics[width=\columnwidth]{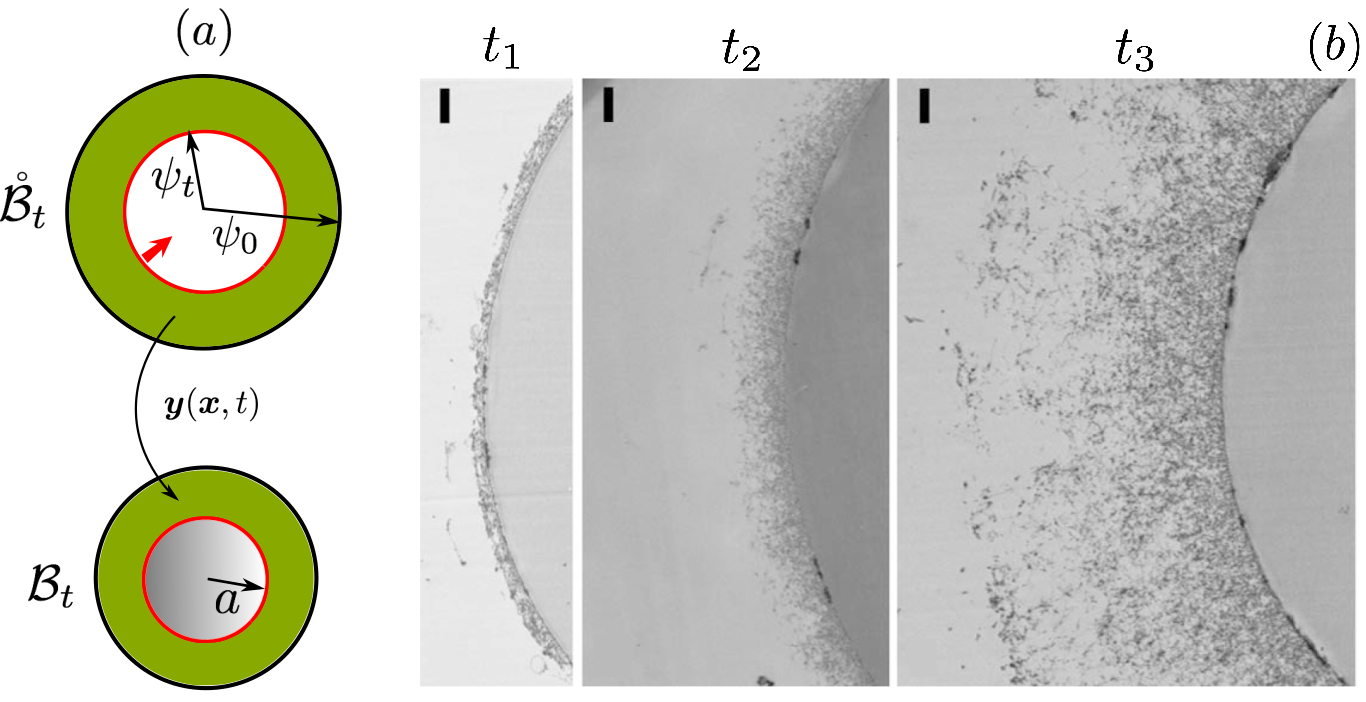}
\caption{\label{actinscheme} Inward accretion over a rigid bead. (a) Schematic 2D representation of the 3D reference and current configurations. The red circle is the trace of the growing surface in both configurations; $\psi_0$ is the initial reference radius and $a$ the current fixed radius of the growing surface.   $(b)$ Electron microscopy images of   an inward  growing actin layer  showing  also outward progression of the external non-growing surface, courtesy  \cite{SykesProst}. }
\end{figure}

\begin{figure}[thb] 
\includegraphics[width=8cm]{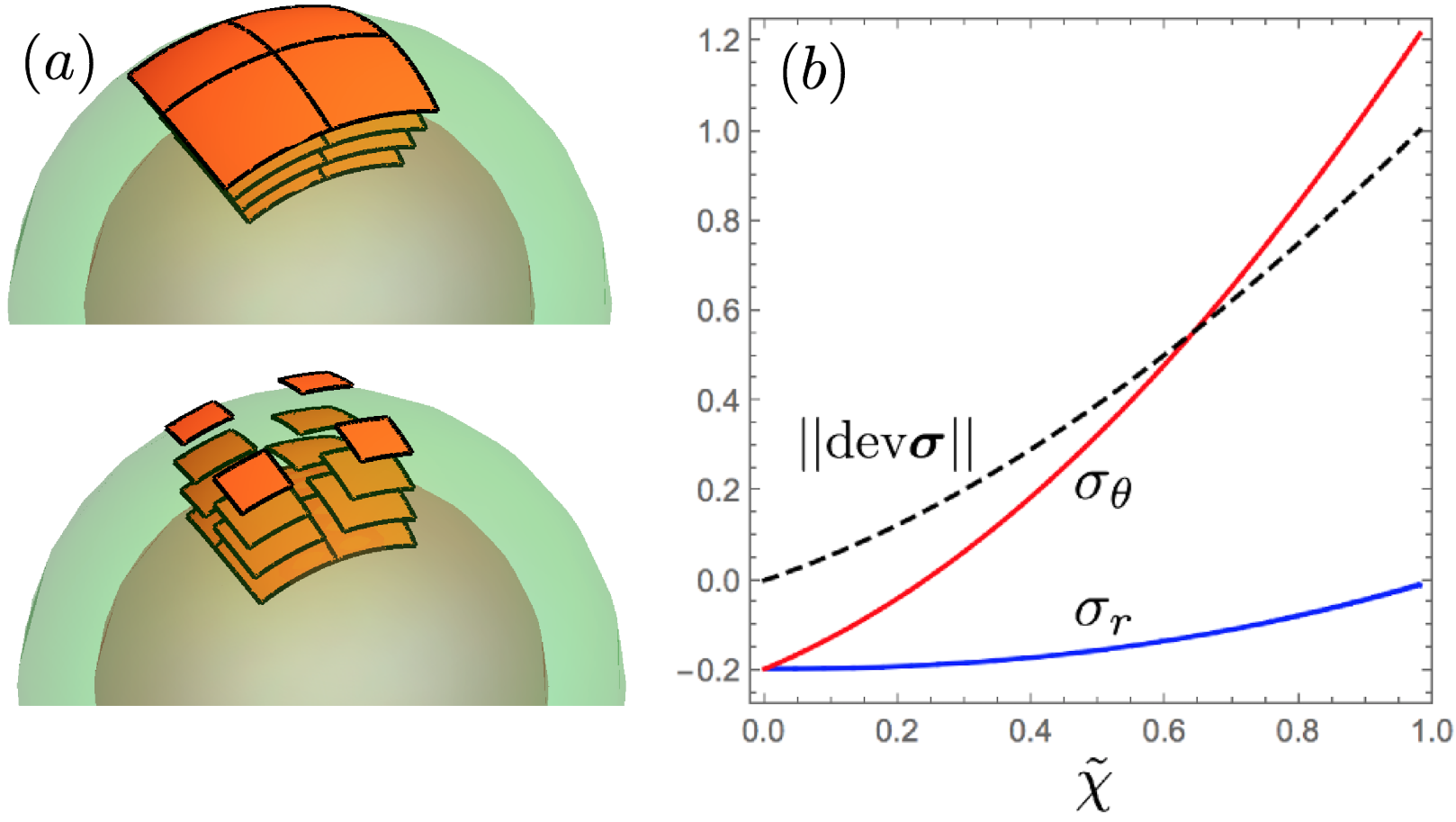}
\caption{\label{actinstress} $(a)$ Schematic representation of the incompatibility developing  in the inward accretion problem. An ideal compatible referential tiling (above) is accreted as   an incompatible tiling  (below) with the gaps  elastically compensated by the growth stresses. The typical  distributions of the   radial (blue), hoop (red) and deviatoric (dashed)  components of these stresses are  shown  in $(b)$. Here $\tilde{\chi}=(\chi-a)/(\psi_0-a)$ is a dimensionless radius in the current configuration. The parameters are: $a=1$, $\psi_0=9$, $\psi_T=4.5$, $ \dot m /(\bar{\varrho}\alpha)=0.05$, $E=5\cdot 10^{-3}$Pa and $\nu=0.3$.}
\end{figure}

\subsection{Inward growth  on a rigid bead\label{secC}} 

Consider now   accretion on a rigid spherical surface, mimicking  the process of inwards actin polymerization stimulated biochemically on a specially treated spherical bead \cite{Tomassetti,Misbah2016,SykesProst}, see Fig.\ref{actinscheme}b. A similar process with  cylindrical symmetry  would be the  growth of  a tree  where new mass is deposited between the existing trunk and the bark \cite{Archer,GuptaTrees}. We use this example to highlight  the essential role of geometrical  nonlinearity because the linearized setting of such a problem is not even meaningful.

Indeed, in our problem  the reference configuration   $\mathring\Body_t= \left\{\bx\,|\,\psi_t\leq||\bx||\leq\psi_0 \right\}$ and the actual  configuration  $\Body_t=\left\{\by\,|\, \chi(\psi_t,t)\leq||\by||\leq \chi(\psi_0,t)\right\}$ are fundamentally different (see Fig.\ref{actinscheme}a), because the current radius of the growing surface is fixed $\chi(\psi_t,t)= a$ while its reference radius $\psi_t$ continuously  evolves. As a result, the reference and actual domains cannot coincide, which is the natural starting assumption of any geometrically linear elasticity theory.

In physical terms, we assume    that the growth process starts at  a surface of a rigid spherical  bead  and that the  arriving  material is being continuously  ``squeezed'' between the emerging  grown body and the original rigid surface  (the actual mechanism  of   mass delivery is obviously disregarded). While  the growing body is expanding away from the bead and its external radius $\chi(\psi_0,t)$ is an increasing function of time,   the  reference radius of the growth  surface  $\psi_t$ is   a decreasing function of  time. 

To determine both, the current state of stress and the two unknown components $(\gamma_r(r),\gamma_{\theta}(r))$ of the reference metric, we impose   three boundary  conditions on the growth surface
\beq
\left\{
\begin{array}{lll}
\bar\s_r(\psi_t) = \bar\s_{\theta}(\psi_t) = -p(\psi_t) \\
\bar{\chi}(\psi_t) = a
\end{array}
\right.
\eeq
The first of this conditions states  that the attachment stress is hydrostatic  with time dependent  pressure  which is controlled externally. In deviation from the previous example,  we assume that the pressure control is not direct but is rather an outcome of the control  of the Eulerian velocity of the arriving material $\bar{\dot{\chi}}(\psi_t) = \dot m /(\bar{\varrho}\alpha)$.
Essentially this means the control of   a volumetric   inflow   rate which  in the context actin polymerization appear to be more realistic   than the    full control of the attachment stress   \cite{Tomassetti,Misbah2016}.  On the exterior surface of the growing body we assume the no-tractions condition, $ s_r(\psi_0,t)=0$.

\begin{figure}[thb] 
\includegraphics[width=7.5cm]{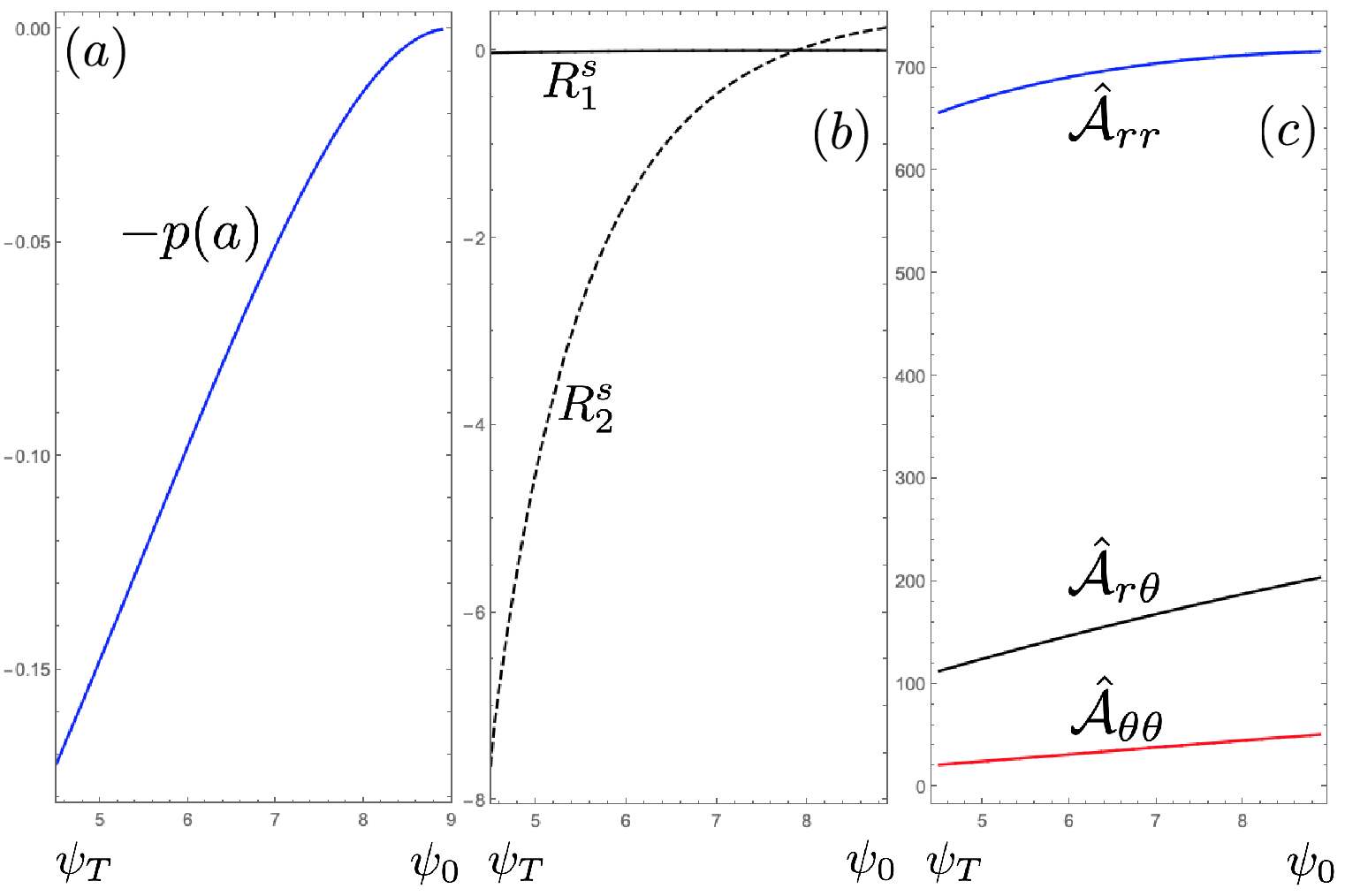}
\caption{\label{Misbah_plots} Inward accretion over a rigid surface. $(a)$ Pressure built-up in the process versus the placement of the referential growing surface, displacing from $\psi_0$ to $\psi_T$. $(b)$ The final distribution of the two components of Ricci curvature and $(c)$ the rescaled elastic moduli in the referential domain $r\in(\psi_T,\psi_0)$. Here, $\hat{\mathcal{A}}_{\alpha\beta}=\tilde{\mathcal{A}}_{\alpha\beta}/\mathcal{A}^0_{\alpha\beta}$, where $\mathcal{A}^0_{\alpha\beta}$ are the moduli \eqref{AAA} evaluated at $s_r=s_{\theta}=0, \lambda_r=\lambda_{\theta}=1$.}
\end{figure}

The succession of incremental problems can be solved numerically  and we illustrate the final  stress distribution  in Fig.\ref{actinstress}b.  Note that the deviatoric stress is maximal at the external (non-growing) surface of the body which,  in principle, should lead to surface instabilities \cite{DeSimone}. The parameters were tuned to match  the numerical  results obtained for a growing network of actin rods,  biased to polymerize on contact with a spherical bead \cite{Misbah2016}. The   pressure build up at the rigid surface, the final distribution of the  components of the Ricci tensor and the residual inhomogeneity  of the elastic moduli, are shown in Fig.\ref{Misbah_plots}. 
 
\subsection{Growth induced material instabilities}  

The aim of our last example is to show that  incompatible mass accretion can lead to material instabilities. Since the letter are not related  directly to geometrical effects,  this  example  stresses the importance of physical  nonlinearity in surface growth problems.  

Suppose that the reference configuration  is given by $\mathring\Body_t= \left\{\bx\,|\,\psi_0\leq||\bx||\leq\psi_t\right\}$ where  now it is the outward surface of a hollow sphere that is  growing.  The current configuration is $\Body_t=\left\{\by\,|\,\chi(\psi_0,t)\leq||\by||\leq\bar\chi(\psi_t)\right\}$,  and we assume again that    the   growing surface is constrained    by a rigid wall $ \chi(\psi_t, t ) = a$. In other wards, we assume that the mass is continuously ``squeezed'' between the existing surface of the body and the rigid spherical cavity, see Fig.\ref{cavitySE}c.  Once again, the actual and the reference  configurations can not coincide at any time  which makes the geometrical linearization of the problem  meaningless. 
\begin{figure}[thb] 
\includegraphics[width=7.5cm]{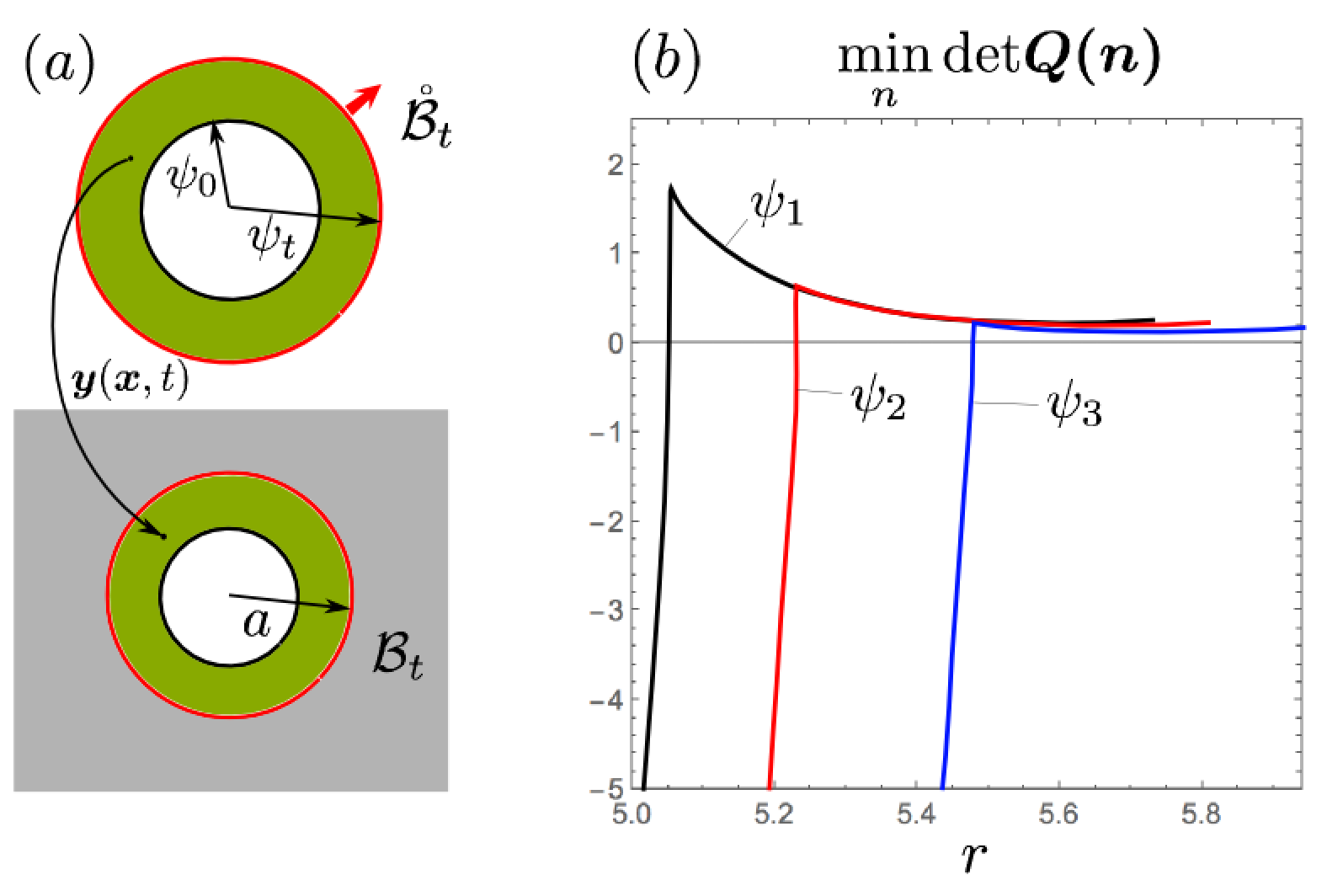}
\caption{\label{cavitySE} $(a)$ Schematic representation of the inward  accretion inside a rigid spherical cavity. The red circle is the trace of the growing surface in the current and reference configurations, the red arrow indicates the growing direction. $(b)$ Distribution of the determinant of the acoustic tensor for three consecutive placements of the growing surface. For $\psi_t>\psi^*=5.82$, strong ellipticity is lost at the internal (non-growing) radius $r^*=A$. Black, blue and red curves describe the determinant for $\psi_1=5.85$, $\psi_2=5.94$ and $\psi_3=6.1$, respectively. In the inset, the plot of the determinant in a wider range, showing the full blown behavior in the negative region. Here $E=1$Pa, $a=3$, $A=\psi_0=5$, $\mathring{D}=1$ and $\dot m/(\bar{\varrho}\alpha) =0.2$.}
\end{figure} 

On the growth surface we maintain the same conditions as in the previous example 
\beq
\left\{
\begin{array}{lll}
\bar\s_r(\psi_t) = \bar\s_{\theta}(\psi_t) = -p(\psi_t) \\
\bar{\chi}(\psi_t) = a,
\end{array}
\right.
\eeq
 and again, instead of prescribing pressure directly,  we assume that the growth is controlled through the Eulerian velocity of the arriving material $\bar{\dot{\chi}}(\psi_t) = - \dot m/(\bar{\varrho}\alpha)$.  Also, similarly to the case of inward growth, we  assume zero tractions condition on the internal surface,  $ s_r(\psi_0,t)=0$. 
 
 \begin{figure}[thb] 
\includegraphics[width=8.5cm]{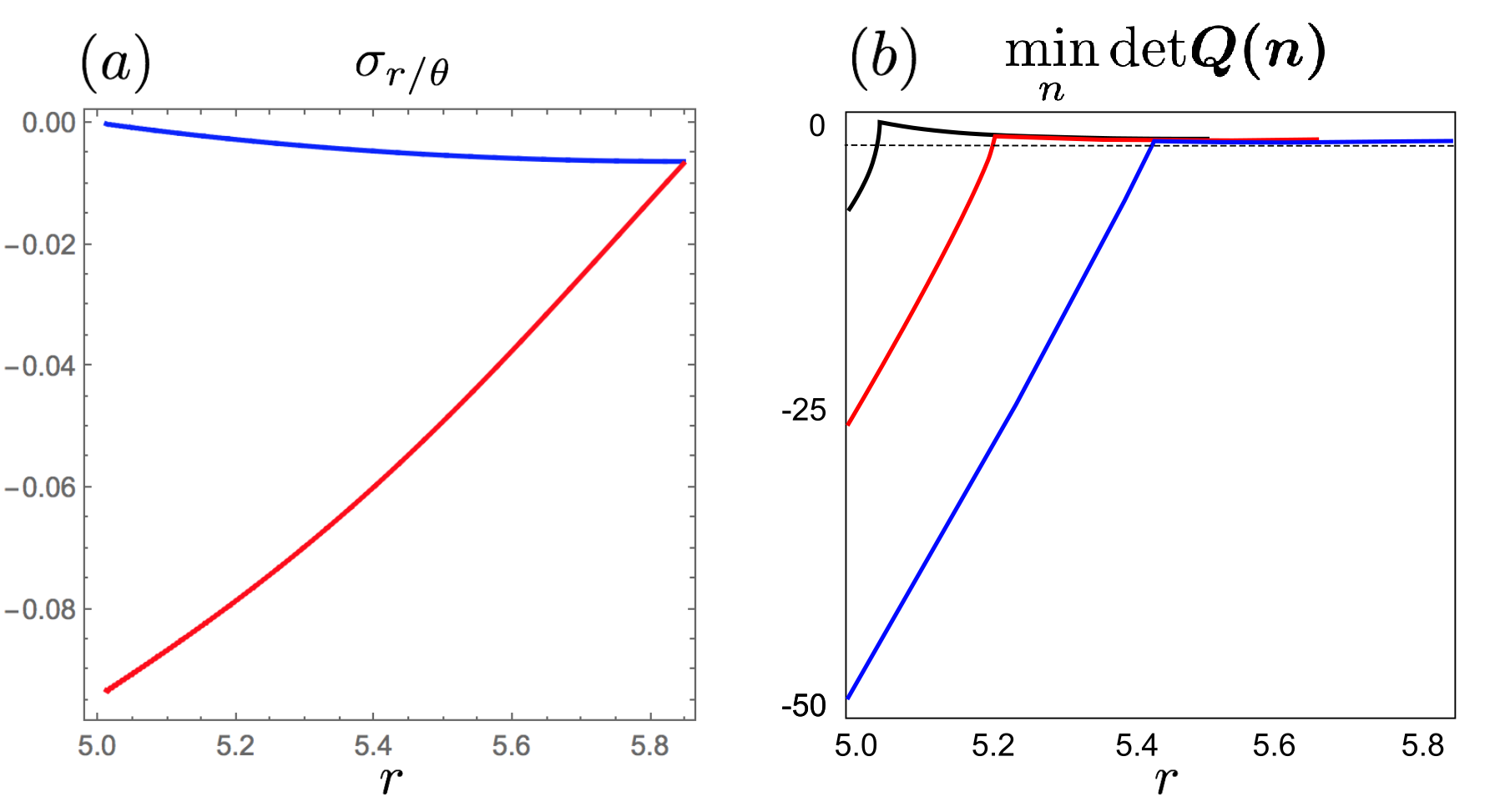}
\caption{\label{cavitySE} Stress distribution in the sphere for $\psi=5.85$. $(b)$ The minimized determinant of the acoustic tensor in a wider range, showing the full blown behavior in the negative region.}
\end{figure} 

The ensuing sequence of incremental problems can be solved numerically and our focus now will be on the  material stability of the grown body. More specifically, we consider the case of linear strain tensor $\beps_e^{\text{lin}}$   (Kirchhoff-St.Venant material)  and check whether the strong ellipticity condition \cite{KnopsPayne} may be violated during accretion.

We recall that the strong ellipticity condition requires that 
$
\mathbfcal{A} (\bm\otimes\mathring \bk)\cdot(\bm\otimes\mathring \bk)>0
$
where $\mathbfcal{A}=\partial\S/\partial\F$ is the  instantaneous elasticity tensor, and $\bm,\mathring \bk$ are  unit vectors. To ensure this condition it is sufficient to check  that  the acoustic tensor $\Q(\bm)$, whose action on a   vector $\mathring \ba$ is defined by  $\Q(\bm)\mathring \ba=\mathbfcal{A}(\mathring \ba\otimes\bm)\bm$, is positive definite \cite{GurtinLinearTh}. To locate the domain of potential material instability we must then search for  the possibility that the determinant $\det\Q(\bm)$ is negative for some  $\bm$  \cite{Ogden}.    

The results of a typical numerical experiment  are shown in Fig.\ref{cavitySE}, where  $\psi_t$ is  the increasing radius of the  growth surface. Before a  critical radius of the growth surface is reached, strong ellipticity is ensured on the whole domain. At the critical threshold the strong ellipticity is first lost at the non-growing (traction free) surface.  As the growth surface advances, the domain where the strong ellipticity is lost,  spreads  behind getting progressively closer to  the growth surface. It should be mentioned, however,  that while  strong ellipticity may be formally lost, our incremental equilibrium problems are constrained  to maintain  radial symmetry. For this reason, our numerical tests do not show instability even after the unconstrained problem  becomes ill-posed.

\section{Conclusions\label{Sec8}}

A large-strain theory of incompatible surface growth is formulated here in terms of the new type of boundary conditions. These conditions are prescribed on a \emph{free surface} which can move along the reference coordinates.  The  novelty of this formulation is in the account of  geometric and constitutive nonlinearities of an elastic body, which are both necessary to adequately capture finite rotations and finite stretches. The theory is presented  in a general 3D setting, with the main focus on path dependence of the incompatible accretion process,  and its goal is to develop the means to \emph{control} the acquired incompatibility by \emph{tailoring} the deposition protocol. 
 
We showed that if both, stresses and displacements, can be independently controlled at the growing surface,  one can ensure point-wise manipulation of the resulting reference metric which  keeps a detailed memory of the deposition process. The relation between the embedded   Ricci tensor of the reference metric and the  time dependence of the control parameters  was shown to be   nonlinear and nonlocal.  It was shown to emerge from a solution of a one parametric family of incremental equilibrium problems for  linear elastic bodies with evolving geometry and  varying elastic inhomogeneity.

The problem simplifies in the case of small deformations when geometry decouples from elasticity, which becomes linear with  elastic moduli both  homogeneous and time independent. We showed that even in this case the incremental elastic problem is characterized by nonstandard boundary conditions. However, the acquired incompatibility can be then expressed analytically in terms of the incremental strain rates which simplifies considerably the dependence  of the deposited reference metric on the surface controls. The main shortcoming of the linearized theory is its inability to deal with  kinematically confined growth, and to account for the possibility of elastic instabilities induced by the growth process.

We illustrated the general theory by a  series of examples emphasizing the role of finite strains in the  surface growth of soft solids. Our examples  highlight the inherent path dependence of the incompatible  growth  and emphasize the effects of  geometrical and physical nonlinearities in ensuring   particular outcomes of the \emph{physically realizable}  growth protocols. Through these examples we showed  that geometrical frustration developing during deposition can be indeed fine-tuned and that  a particular behavior of a system in physiological  or  industrial conditions can be engineered by embedding into the material a particular incompatibility. The proposed general theoretical framework allowing one to handle  such  \emph{information-rich} solids can be used to guide bio-mimetic design and additive manufacturing.   

The fact that one can engineer  the incompatibility in  a manufactured solid body and regulate in this way the associated  distribution of residual stresses is of crucial  importance for the understanding of biological processes  where solids  are typically functionally pre-stressed. Our analysis also highlights the possibility to manufacture  non-Euclidean elastic solids artificially, not only with  a tailored pre-stress, but also with a  particular stiffness distribution. Such solids can be designed to undergo  specific elastic instabilities and to exhibit specific patterning  in  the technologically relevant conditions \cite{Gao12,Sharon11,Bangham,Roman,Burgueno,2Dprinting}.

\section{Acknowledgments} The authors thank V.Balbi, D.Barilari, R.Kupferman, E.Sharon, B.Shoikhet and G.Tomassetti for helpful discussions. G.Z. acknowledges the PMMH-ESPCI for hospitality during 2018, the sabbatical support from NUI-Galway and the support of the Italian National Group of Mathematical Physics (INDAM-GNFM). L.T. was supported by the French government  under the  Grant No. ANR-10-IDEX-0001-02 PSL.


\begin{thebibliography}{100}

\bibitem{Archer} Archer R.R., Growth Stresses and Strains in Trees, Springer Series in Wood Science, Springer-Verlag Berlin Heidelberg (1987). 

\bibitem{Dumais2001} Dumais J., Kwiatkowska D., Analysis of surface growth in shoot apices, The Plant J., 31(2), 229--241 (2001). 

\bibitem{Dafalias} Dafalias Y.F., Pitouras Z., Stress field in actin gel growing on spherical substrate, Biomech. Model. Mechanobiol. 8, 9--24 (2009). 

\bibitem{johnpre14} John K., Caillere D., Misbah C.,  Spontaneous polarization in an interfacial growth model for actin filament networks with a rigorous mechanochemical coupling, Phys. Rev. E 90, 052706 (2014). 

\bibitem{GoodmanSlopes} Goodman, L. E., Brown, C. B.  Dead Load Stresses and the Instability of Slopes, J. Soil Mech. Found. Div., Proc. A.S.C.E., 89, SM3, 103-134, (1963). 

\bibitem{Labuz} Bentler J.G., Labuz J.F., Performance of a Cantilever Retaining Wall, J. Geotech. Geoenviron. Eng.132:1062-1070 (2006). 

\bibitem{Kadish2005} Kadish J., Barber J.R., Washabaugh P.D., Stresses in rotating spheres grown by accretion, Int.J.Sol.Struct. 42, 5322--5334 (2005). 

\bibitem{Wildeman} Wildeman S., Sterl S., Sun C., Loshe D., Fast dynamics of water droplets freezing from the outside in, Phys. Rev. Lett., 118, 084101 (2017). 

\bibitem{Fink} Gumennik A., Levy E.C., Grena B., Hou C., Rein M., Abouraddy A.F., Joannopoulos J.D., Fink Y., Confined in-fiber solidification and structural control of silicon and silicon--germanium microparticles, Proc.Nat.Amer.Soc., 114(28), 7240--7245 (2017). 

\bibitem{Schwerdtfeger} Schwerdtfeger K., Sato M., Tacke K.H., Stress formation in solidifying bodies. Metallurgical and Materials Transactions B, 29,5,1057--1068 (1998).

\bibitem{Ge} Ge, Qi, et al., Multimaterial 4D Printing with Tailorable Shape Memory Polymers, Scientific Reports 6, 31110 (2016). 

\bibitem{LindNature16} Lind J.U. et al, Instrumented cardiac microphysiological devices via multimaterial three-dimensional printing, Nature Mat. 16, 303-308 (2017). 

\bibitem{Skalak} Skalak R., Hoger A., Kinematics of surface growth. J. Math. Biol. 35, 869--907 (1997). 

\bibitem{Epstein} Epstein M., Kinetics of boundary growth, Mech. Res. Comm. 37(5), 453--457 (2010).

\bibitem{DiCarloSurf} DiCarlo A. Surface and Bulk Growth Unified. In: Steinmann P., Maugin G.A. (eds) Mechanics of Material Forces. Advances in Mechanics and Mathematics, vol 11. Springer, Boston, MA (2005). 

\bibitem{MauginCiarletta} Ciarletta P., Preziosi L., Maugin G.A., Mechanobiology of interfacial growth, J. Mech. Phys. Solids, 61, 852--872 (2013). 

\bibitem{Ganghoffer2010} Ganghoffer J.-F., Mechanical modeling of growth considering domain variation -- Part II:
Volumetric and surface growth involving Eshelby tensors, J.Mech.Phys.Solids 58, 1434--1459 (2010). 

\bibitem{Tomassetti} Tomassetti G., Cohen T.,  Abeyaratne R., Steady accretion of an elastic body on a hard spherical surface and the notion of a four-dimensional reference space, J. Mech. Phys. Solids, 96, 333--352 (2016).

\bibitem{MoultonGoriely} Moulton D.E., Goriely A., Chirat R., Mechanical growth and morphogenesis of seashells, J. Theoret. Biology 311, 69--79 (2012). 

\bibitem{Jenkins} Buskohl P.R., Butcher J.T., Jenkins J.T., The influence of external free energy and homeostasis on growth and shape change, J.Mech.Phys.Solids 64, 338--350 (2014). 

\bibitem{kondo49} Kondo, K., A proposal of a new theory concerning the yielding of materials based on Riemannian geometry, I.J Soc. Appl. Mech. Japan, 2(11), 123-128 (1949). 

\bibitem{Efrati} Efrati E., Sharon E., Kupferman R., The metric description of elasticity in residually stressed soft materials, Soft Matter 9, 8187 (2013). 

\bibitem{RazPNAS} Moshe M., Levin I., Aharoni H., Kupferman R., Sharon E., Geometry and mechanics of two-dimensional defects in amorphous materials, Proc.Nat.Acad.Sci. 112(35), 10873--10878 (2015). 

\bibitem{Sharon} Klein, Y., Efrati, E.,Sharon, E., Shaping of Elastic Sheets by Prescription of Non-Euclidean Metrics. Science, 315(5815), 1116--1120 (2007).

\bibitem{GorielyPlants} Geitmann A. et al., Actuators Acting without Actin, Cell 166(1) 30,15-17 (2016). 

\bibitem{Hossain} Hossain A.B., Weiss J., Assessing residual stress development and stress relaxation in restrained concrete ring specimens, Cement Concrete Comp. 26(5) 531-540 (2004). 

\bibitem{Martley} Martley J.F., Theoretical calculations of the pressure distribution on the basal section of a tree, Forestry 2, 69-72 (1928). 

\bibitem{Arutyunyan} Arutyunyan N. Kh., Metlov V. V.,Izv. Akad. Nauk SSSR, Mekh. Tverd. Tela 4, 142--152 (1983).

\bibitem{Southwell} Southwell R., Introduction to the Theory of Elasticity for Engineers and Physicists, Oxford University Press (1941). 

\bibitem{Rashba} Rashba, E. I., Stress determination in bulks due to own weight taking into account the construction sequence,  Proc. Inst. Struct. Mech. Acad. Sci. Ukrainian SSR, 18, 23-27 (1953).

\bibitem{Trincher} Trincher V.K., Izv. AN SSSR. Mekhanika Tverdogo Tela, 19(2) 119-124 (1984). 

\bibitem{Naumov} Naumov V. E., Mechanics of growing deformable solids: a review, J. Eng. Mech.,120, 207-220 (1994). 

\bibitem{Manz14} Manzhirov A.V., Lychev S.A., Mathematical modeling of additive manufacturing technologies, Proc.World.Congress.Engng. (WCE) Vol.II, ISBN: 978-988-19253-5-0 (2014). 

\bibitem{Drozdov98} Drozdov A.D., Viscoelastic Structures: Mechanics of Growth and Aging, AcademicPress, NewYork (1998). 

\bibitem{Lychev17} Lychev S.A. Geometric Aspects of the Theory of Incompatible Deformations in Growing Solids. In: Altenbach H., Goldstein R., Murashkin E. (eds) Mechanics for Materials and Technologies. Advanced Structured Materials, vol 46. Springer, Cham (2017). 

\bibitem{Goodman} Brown C.B., Goodman L.E., Gravitational Stresses in Accreted Bodies, Proc.Royal Soc. London Sez.A, Math. and Phys. Sci., 276(1367), 571--576 (1963). 

\bibitem{Fletcher} King W.D., Fletcher N.H., J.Phys.D: Appl.Phys. 6(18), 21--57  (1973). 

\bibitem{Zabaras} Zabaras N., Liu S., A theory for small deformation analysis of growing bodies with an application to the winding of magnetic tape packs, Acta Mech. 111, 95--110 (1995). 

\bibitem{Gambarotta} Bacigalupo A., Gambarotta L., Mechanics Based Design of Structures and Machines 40, 163--184 (2012). 

\bibitem{SozioYavari} Sozio F., Yavari A., Nonlinear mechanics of surface growth for cylindrical and spherical elastic bodies, J. Mech. Phys. Solids 98, 12-48 (2017). 

\bibitem{GuptaTrees} Swain D, Gupta A., Biological growth in bodies with incoherent interfaces. Proc. R. Soc. A 474: 20170716 (2018). 

\bibitem{GanghofferGoda} Ganghoffer J-F., Goda I., Multiscale Biomechanics, Chap.9 - Multiscale Aspects of Bone Internal and External Remodeling, Ed. J-F.Ganghoffer, ISBN 9781785482083, Elsevier (2018). 

\bibitem{Papadopoulos} Hodge N. Papadopoulos P.,  A continuum theory of surface growth, Proc. R. Soc. A 466, 3135-3152 (2010). 

\bibitem{Goriely19} S Rudraraju S., Moulton D.E., Chirat R., Goriely A., Garikipati K., A computational framework for the morpho-elastic development of molluskan shells by surface and volume growth
arXiv preprint arXiv:1901.00497 (2019).

\bibitem{Bulging} Weickenmeier J., Saez P., Butler C.A.M., Young P.G., Goriely A., Kuhl E.,  Bulging brains, J.Elast.129(1-2), 197-212 (2017). 

\bibitem{ZTprl} Zurlo G., Truskinovsky L., Printing Non-Euclidean Solids,  Phys. Rev. Lett., 119, 048001 (2017). 

\bibitem{ZTMaugin} Zurlo G., Truskinovsky L., Inelastic surface growth, Mech. Res. Commun. 93
174--179 (2018). 

\bibitem{Misbah2016} John K., Stoter T., Misbah C., A variational approach to the growth dynamics of pre-stressed actin filament networks, J. Phys.: Condens. Matter 28 375101 (2016). 

\bibitem{MacKintosh} Broedersz C.P., MacKintosh F.C., Modeling semiflexible polymer networks,  Rev. Mod. Phys. 86 (3), 995--1036 (2014). 

\bibitem{Nardinocchi}  Minozzi, M., Nardinocchi, P., Teresi, L., Varano, V.  Growth-induced compatible strains Math.Mech.Sol., 22(1), 62--71 (2017).

\bibitem{Teresi} Lucantonio A., Nardinocchi P., Teresi L., Transient analysis of swelling-induced large deformations in polymer gels, J. Mech. Phys. Solids 61 205--218 (2013). 

\bibitem{Ball} Ball J., Convexity conditions and existence theorems in nonlinear elasticity, Arch.Rat.Mech.Anal., 63(4), 337--403 (1976). 

\bibitem{Vanel} Vanel L., Howell D., Clark D., Behringer R. P., Clement E., Memories in sand: Experimental tests of construction history on stress distributions under sandpiles, Phys. Rev. E 60, R5040(R) (1999).

\bibitem{Kuhnel} K$\ddot{\text{u}}$hnel W., Differential Geometry, Curves - Surfaces - Manifolds, AMS (2006).

\bibitem{Ciarlet} Ciarlet P.G., An Introduction to Differential Geometry with Applications to Elasticity, Springer (2005).

\bibitem{Davini} Davini C., Some remarks on the continuum theory of defects in solids, Int.J.Sol.Str. 38, 1169--1182 (2001).   

\bibitem{Hsu68} Hsu, F.H., The influences of mechanical loads on the form of a growing elastic body.
J. Biomech. 1(4), 303--311 (1968). 

\bibitem{Danescu2018} Danescu, A., Regreny, P., Cremillieu, P., Leclercq, J. L., Fabrication of self-rolling geodesic objects and photonic crystal tubes, Nanotechnology, 29(28), 285301 (2018). 

\bibitem{Skalak82} Skalak R., Growth as a finite displacement field, D. E. Carlson et al. (eds.), Proceedings of the IUTAM Symposium on Finite Elasticity, Martinus Nijhoff Publishers, The Hague, 347-355 (1982). 

\bibitem{NoteBar} For an arbitrary scalar, vectorial or tensorial field $\varphi$, we denote by $\bar\varphi(\bx):=\varphi(\bx,\mathring\vartheta(\bx))$.  

\bibitem{GreenNaghdi} Green A.E., Naghdi P.M., A General Theory of an Elastic-Plastic Continuum, Arch. Rational Mech. Anal.18, 251--281 (1965). 

\bibitem{Bruhns} Bruhns, O. T., Xiao, H., Meyers, A., Constitutive inequalities for an isotropic elastic strain energy function based on Hencky's logarithmic strain tensor. Proc. Roy. Soc. London A 457, 2207--2226 (2001). 

\bibitem{Auricchio} Arghavani J., Auricchio F., Naghdabadi R., A finite strain kinematic hardening constitutive model based on Hencky strain: General framework, solution algorithm and application
to shape memory alloys, Int. J. Plasticity 27, 940-961 (2011). 

\bibitem{Anand} Anand L., On H. Hencky's approximate strain-energy function for moderate deformations, J. Appl. Mech. 46, 78-82 (1979).

\bibitem{VanGoethem} Maggiani G.B., Scala R., Van Goethem N., A compatible-incompatible decomposition of symmetric tensors in Lp with application to elasticity, Math.Meth.Appl.Sci. 38, 5217--5230 (2015).

\bibitem{FosdickEdelstein} Edelstein W.S., Fosdick R.L., A note on non-uniqueness in linear elasticity theory, ZAMP 19(6), 906--912 (1968). 

\bibitem{RazPRE} Moshe M., Sharon E., Kupferman R., Elastic interactions between two-dimensional geometric defects, Phys.Rev. E 92, 062403 (2015). 

\bibitem{RazARMA} Kupferman R, Moshe M, Solomon J.P., Metric description of singular defects in
isotropic materials, Arch. Ration. Mech. Anal. 216:1009--1047 (2014). 

\bibitem{Danescu} Danescu A., Chevalier C., Grenet G., Regreny Ph., Letartre X., Leclercq J. L., Spherical curves design for micro-origami using intrinsic stress relaxation, Appl. Phys.Lett. 102, 123111 (2013).

\bibitem{BenAmarConicalDefects} Mueller M.M., Ben Amar M., Guven J., Conical Defects in Growing Sheets, Phys. Rev. Lett. 101, 156104 (2008). 

\bibitem{Warner} Modes C.D., Bhattacharya K., Warner M., Gaussian curvature from flat elastica sheets, Proc. R. Soc. A 467, 1121--1140 (2011). 

\bibitem{HanHong} Han Q., Hong J.X., Isometric Embedding of Riemannian Manifolds in Euclidean Spaces, AMS 130, (2006).

\bibitem{Nash} Nash J., C$^1$ Isometric imbeddings, Annals of Mathematics, 60(3), 383--396 (1954). 

\bibitem{SykesProst}  Plastino J., Lelidis I.,Prost J., Sykes C.,  The effect of diffusion, depolymerization and nucleation promoting factors on actin gel growth, Eur Biophys J. 33, 310-320 (2004). 

\bibitem{DeSimone} Cardamone L., Laio A., Torre V., Shahapure R., DeSimone A., Cytoskeletal actin networks in motile cells are critically self-organized systems synchronized by mechanical interactions, Proc. Nat. Acad. Sci.108 (34), 13978-13983 (2011). 

\bibitem{KnopsPayne} Knops R.J., Payne L.E., Uniqueness Theorems in Linear Elasticity, Springer-Verlag New York Heidelberg Berlin (1971).

\bibitem{GurtinLinearTh} Gurtin M.E., The Linear Theory of Elasticity. In: Truesdell C. (eds) Linear Theories of Elasticity and Thermoelasticity. Springer, Berlin, Heidelberg (1973). 

\bibitem{Ogden} Ogden R.W., Non-Linear Elastic Deformations, Dover Publications (1997). 

\bibitem{Gao12} Li, B., Cao, Y. P., Feng, X. Q., Gao, H. Mechanics of morphological instabilities and surface wrinkling in soft materials: a review. Soft Matter, 8(21), 5728-5745 (2012). 

\bibitem{Sharon11} Shahaf A., Efrati E., Kupferman R., Sharon E., Geometry and mechanics in the opening of chiral seed pods, Science 333(6050), 1726-1730 (2011). 

\bibitem{Bangham} Kennaway R., Coen E., Green A., Bangham A. Generation of diverse biological forms through combinatorial interactions between tissue polarity and growth, PLoS Comp. Biology, 7(6), e1002071 (2011).

\bibitem{Roman} Siefert  E., Reyssat E., Bico J., Roman B., Bio-inspired pneumatic shape-morphing elastomers, Nature Materials, 18(1), 24 (2019).

\bibitem{Burgueno} Hu N., Burgueno R., Buckling-induced smart applications: recent advances and trends, Smart Mat. Struct.  24.6, 063001 (2015). 

\bibitem{2Dprinting}  Cafferty B.J., Campbell V.E., Rothemund P., Preston D.J., Ainla A., Fulleringer N., Diaz A.C., Fuentes A.E., Sameoto D., Lewis J.A., Whitesides G.M., Fabricating 3D Structures by Combining 2D Printing and Relaxation of Strain. Advanced Materials Technologies, p.1800299 (2018).

\end{thebibliography}
\end{document}